\documentclass{aa}  

\usepackage{graphicx}
\usepackage{natbib} 
\bibpunct{(}{)}{;}{a}{}{,} % to follow the A&A style

\usepackage[varg]{txfonts}
\newcommand{\kms}{km\,s$^{-1}$}
\newcommand{\um}{$\mu$m}

\newcommand{\cms}{cm$^{-2}$}
\newcommand{\cmc}{cm$^{-3}$}

\newcommand{\lsol}{L$_{\odot}$}

\newcommand{\hcop}{HCO$^{+}$}

\newcommand{\hccop}{H$^{13}$CO$^{+}$}
\newcommand{\hcoop}{HC$^{18}$O$^{+}$}
\newcommand{\hcsp}{HCS$^{+}$}
\newcommand{\hcnhp}{HCNH$^{+}$}
\newcommand{\hocop}{HOCO$^{+}$}
\newcommand{\cop}{CO$^{+}$}
\newcommand{\sop}{SO$^{+}$}
\newcommand{\nnhp}{N$_{2}$H$^{+}$}

\newcommand{\nhcop}{N$_{\rm HCO^{+}}$}
\newcommand{\xhcop}{X$_{\rm HCO^{+}}$}
\newcommand{\nnnhp}{N$_{\rm N_2H^{+}}$}
\newcommand{\xnnhp}{X$_{\rm N_2H^{+}}$}
\newcommand{\nhcsp}{N$_{\rm HCS^{+}}$}
\newcommand{\xhcsp}{X$_{\rm HCS^{+}}$}
\newcommand{\nsop}{N$_{\rm SO^{+}}$}
\newcommand{\xsop}{X$_{\rm SO^{+}}$}
\newcommand{\nhocop}{N$_{\rm HOCO^{+}}$}
\newcommand{\xhocop}{X$_{\rm HOCO^{+}}$}
\newcommand{\denshh}{n$_{\rm H_{2}}$}
\newcommand{\nhh}{N$_{\rm H_{2}}$}
\newcommand{\xcoo}{X$_{\rm CO_2}$}
\newcommand{\xcs}{X$_{\rm CS}$}
\newcommand{\xocs}{X$_{\rm OCS}$}
\newcommand{\xhhs}{X$_{\rm H_{2}S}$}

\newcommand{\hhhp}{H$_{3}$$^{+}$}
\newcommand{\hhp}{H$_{2}$$^{+}$}
\newcommand{\splus}{S$^{+}$}

\newcommand{\hh}{H$_{2}$}

\newcommand{\ho}{H$_{2}$O}

\newcommand{\coo}{CO$_{2}$}

\begin{document}
   \title{Molecular ions in the protostellar shock L1157-B1}

   \author{L. Podio
          \inst{1,2}
          \and
          B. Lefloch
          \inst{1}
          \and
          C. Ceccarelli
          \inst{1}
          \and
          C. Codella
          \inst{2}
          \and
          R. Bachiller
          \inst{3}
          }

   \institute{
     Institut de Plan\'etologie et d'Astrophysique de Grenoble, 414, Rue de la Piscine, 38400 St-Martin d'H\`eres, France  \\
     \email{lpodio@arcetri.astro.it}
   \and 
     INAF - Osservatorio Astrofisico di Arcetri, Largo E. Fermi 5, 50125 Firenze, Italy
   \and 
     IGN Observatorio Astron\'omico Nacional, Apartado 1143, 28800 Alcal\'a de Henares, Spain
             }

   \date{Received ; accepted }

% \abstract{}{}{}{}{} 
% 5 {} token are mandatory
 
  \abstract
{}
%The shocks occurring when the material ejected by a young star impacts the surrounding medium strongly affect the gas composition and physical conditions.
%, for example they enhance the abundances of molecules which are usually locked onto dust grains. 
%Very little is known about their effect on the chemistry of molecular ions.
   {
%We search for molecular ions in the B1 shock along the L1157 outflow and we investigate their chemical evolution.
%if the detected lines are probing pre-shock conditions in the cloud, or if instead their abundance is enhanced in the shock.
%We investigate the chemistry of molecular ions in shocks.
In this work, we perform a complete census of molecular ions with an abundance larger than $\sim 10^{-10}$ in the protostellar shock L1157-B1. This allows us to study the ionization structure and chemistry of the shock.}
   {An unbiased high-sensitivity survey of L1157-B1 performed with the IRAM-30m and {\it Herschel}/HIFI as part of the CHESS and ASAI large programs allows searching for molecular ions emission.
%a census of molecular ions in the  B1 shock along the L1157 outflow. Different velocity components are disentangled in the line profiles. 
Then, by means of a radiative transfer code in the Large Velocity Gradient approximation, the gas physical conditions and fractional abundances of molecular ions are derived. The latter are compared with estimates of steady-state abundances in the cloud and their evolution in the shock
%due to density and temperature enhancement (\denshh~$=10^5$~\cmc, T~$=70$~K) and to the release of molecules off dust mantles, 
calculated with the chemical model Astrochem. }
%and compared with observations.}
   {We detect emission from \hcop, \hccop, \nnhp, \hcsp, and, for the first time in a shock, from \hocop, and \sop.
The bulk of the emission peaks at blueshifted velocity, $\sim 0.5-3$~\kms\ with respect to systemic, has a width of $\sim 4-8$~\kms\ and is associated with the outflow cavities (T$_{\rm kin}\sim20-70$~K,  \denshh~$\sim10^5$~\cmc).
A high velocity component, up to $-40$~\kms, associated with the primary jet is detected in the \hcop~1--0 line.
Observed \hcop\ and \nnhp\ abundances (\xhcop~$\sim0.7-3~10^{-8}$, \xnnhp~$\sim0.4-8~10^{-9}$) are in agreement with steady-state abundances in the cloud and with their evolution in the compressed and heated gas in the shock for cosmic rays ionization rate $\zeta=3~10^{-16}$~s$^{-1}$. %, suggesting that \hcop\ and \nnhp\ are probing pre-shock chemistry.
\hocop, \sop, and \hcsp\ observed abundances (\xhocop~$\sim10^{-9}$, \xsop~$\sim8~10^{-10}$, \xhcsp~$\sim3-7~10^{-10}$), instead, are 1--2 orders of magnitude larger than predicted in the cloud; on the other hand they are strongly enhanced on timescales shorter than the shock age ($\sim$2000 years) if \coo, S or H$_2$S, and OCS are sputtered off the dust grains in the shock. }
{The performed analysis indicates that \hcop\ and \nnhp\ are a fossil record of pre-shock gas in the outflow cavity, while \hocop, \sop, and \hcsp\ are effective shock tracers and can be used to infer the amount of \coo\ and sulphur-bearing species released from dust mantles in the shock. The observed \hcsp\ (and CS) abundance indicates that OCS should be one of the main sulphur carrier on grain mantles. However, the OCS abundance required to fit the observations is 1--2 orders of magnitude larger than observed. Laboratory experiments are required to measure the reactions rates involving these species and to fully understand the chemistry of sulphur-bearing species.}


   \maketitle
%
%________________________________________________________________

\section{Introduction}

The gravitational infall from which a new star is formed is accompanied by the ejection of highly supersonic jets.
The shocks produced when the ejected material impacts on the high-density surrounding medium rapidly heat and compress the gas triggering several microscopic processes such as molecular dissociation, gas ionization, endothermic chemical reactions, ice sublimation, and dust grain disruption.
It is well know that these processes produce a significant enhancement of the abundance of molecules which are evaporated or sputtered from dust grains, such as H$_2$CO, CH$_3$OH, NH$_3$, SiO  \citep[see, e.g., ][]{bachiller97}.
%On the other hand, very little is known about the processing of molecular ions in shocks.

Molecular ions are commonly observed in dense prestellar/protostellar regions \citep[e.g., ][]{caselli98,caselli02c,hogerheijde98} and their chemistry have been studied by means of gas-phase chemical models \citep[e.g., ][]{herbst86,turner99}.
In contrast, there are very few observations of molecular ions in protostellar shocks. 
\hcop\ was detected in a number of outflows and in some cases show a 'butterfly' morphology suggesting excitation in the outflow cavities \citep[e.g., ][]{hogerheijde98,girart99,girart02,lee07,tafalla10,tappe12}.
Besides \hcop, emission in \hcsp\ and very recently in \nnhp\ lines was detected in the B1 shock along the chemically rich L1157 outflow \citep{bachiller97,yamaguchi12,codella13}.

Due to the lack of observations, the chemical evolution of molecular ions in shocks has been poorly investigated. 
On the one hand, molecular ions are expected to be rapidly destroyed by electronic recombination in shocks; on the other hand, they may be enhanced due to the sputtering of dust grains mantles, which injects molecules in the gas phase where they undergo reactions leading to ions \citep[e.g., ][]{neufeld89a,viti02,rawlings04}.
%For example, the abundance of \hcop\ and \nnhp\ is thought to decrease in the post-shocked gas due to the density enhancement which cause more efficient recombination. 
Among the  molecular ions previously observed in shocks, not all of them are due to changes caused by the shock itself.
\citet{codella13} show that the \nnhp\ abundance is expected to decrease in shocks, hence the emission detected in the L1157-B1 shock is a probe of pre-shock chemistry.  
%Moreover, \citet{viti02,viti03} and \citet{rawlings00,rawlings04} show that the abundance of some molecular ions, e.g. \hcop\ and \hcsp\, may be enhanced in strong shock where the gas, enriched by the release of molecules from the icy mantles of dust grains, is irradiated by shock-induced UVs. The produced abundance enhancement takes place on timescales of $\sim$100 years but may be observed on a few thousands years as the outflow continues eroding the cavity interacting with fresh material.
Molecular ions such as \hocop\ and \sop, instead, are believed to be effective tracers of shocked gas, as their abundance is predicted to be very low in the quiescent gas ($\le 10^{-10}$) but could be strongly enhanced in shocks following the release of \coo\ and S-bearing molecules from dust grains mantles 
%and the gas dissociation and ionization (producing \splus) 
\citep[e.g., ][]{minh91,turner92,turner94,deguchi06}.
However, there are no observations of \hocop\ and \sop\ undoubtedly associated with shocks.
%In particular, the \sop\ abundance is predicted to be increased by up to six order of magnitude in dissociative shocks \citep{neufeld89a} and 
%On the other hand, \hocop\ is mainly formed by the reaction between \hhhp\ and \coo, thus 
%previous observations suggest that \hocop\ could be strongly enhanced in shocks following \coo\ release from the icy mantle of dust grains \citep{deguchi06,sakai08}.
%Observational estimates of the abundances of molecular ions may help us to understand their evolution in the cloud and in the shock.

%Moreover, once molecular species are released back to the gas phase, the enriched gas may be irradiated by shock-induced UV radiation, producing molecular ions such as \hcop, H$_3$O$^{+}$, \sop, \cop, \hcsp, and \nnhp\ \citep{viti02,viti03,rawlings04}.   Models predict column densities of the most abundant molecular ions of ..... \citep{viti02}. 
%Observations of molecular ions is crucial to test these predictions and constrain the shock physics.
%However, there are very few observations of molecular ions in outflows.
%While \hcop\ lines were observed in a number of outflows \citep[e.g., ][]{girart99,girart02,bachiller97,lee07,tafalla10,tappe12},  emission from its isotopologues (e.g. \hccop), as well from other ions (\nnhp, \cop, and \sop) was reported only at the source position \citep{bachiller97,bachiller01,stauber06}.

%For the first time, we have obtained a complete census of the molecular ion content  in a protostellar shock, down to a sensitivity of \approx 1 mK.
In this paper, for the first time a complete census of molecular ions in a shock is given.
%we present a comprehensive survey of molecular ions in the millimeter and submillimeter range in the prototypical  protostellar bow-shock L1157-B1. 
%down to a sensitivity of $\sim 1.5$~mK per 1.4 \kms\ interval in the 3~mm band. 
This is obtained by means of a highly-sensitive unbiased spectral survey of the prototypical  protostellar bow-shock L1157-B1  in the millimeter and submillimeter range, executed with the IRAM-30m telescope as part of the ASAI Large Program\footnote{http://www.oan.es/asai} (Astrochemical Surveys At Iram, Lefloch, Bachiller et al. in preparation) and with {\it Herschel}/HIFI as part of the Guaranteed Time Key Project CHESS\footnote{http://chess.obs.ujf-grenoble.fr} (Chemical HErschel Surveys of Star forming regions, \citealt{ceccarelli10}). 
%In addition to \hcop, \nnhp, and \hcsp\ previously identified by \citet{bachiller97}, \citet{yamaguchi12}, and \citet{codella13}, we report for the first time emission by \hocop\ and \sop\ in a protostellar shock. 
The observations are presented in Sect.~\ref{sect:observations}, while Sect.~\ref{sect:l1157} summarizes the main properties of the L1157 outflow and the B1 bow-shock. 
In Sect.~\ref{sect:results} we analyze the line profiles of the detected molecular ions and we perform a Large Velocity Gradient (LVG) analysis to infer the physical conditions of the emitting gas, and the abundance of molecular ions. 
%Then we estimate the abundance and fractional enhancement of the detected molecular ions with respect to the values derived at the source position when available.
Finally, in Sect.~\ref{sect:astrochem} we compare the observed abundances with steady-state abundances in the cloud and their evolution in the shock computed using the chemical code Astrochem.
This allows us to distinguish between molecular ions which are tracing pre-shock chemistry, and those which are a useful probe of the shock properties and chemistry.
Our conclusions are summarized in Sect.~\ref{sect:conclusions}.

\section{Observations and data reduction}
\label{sect:observations}

\subsection{IRAM-30m observations}

The millimeter observations of  L1157-B1 were acquired with the IRAM-30m telescope at Pico Veleta (Spain), as part of the ASAI Large Program (Lefloch, Bachiller et al. in preparation).
The observed position in B1 is at $\alpha_{\rm J2000}$ = 20$^{\rm h}$ 39$^{\rm m}$ 10$\fs$2, $\delta_{\rm J2000}$ = +68$\degr$ 01$\arcmin$ 10$\farcs$5. 
The survey, obtained during several runs in 2011 and 2012, covers the spectral bands at 3 mm (80--116 GHz), 2 mm (128--173 GHz), 1.3 mm (200--320 GHz), and 0.8 mm (328--350 GHz).  
The observations were carried out in Wobbler Switching Mode, with a throw of 3\arcmin, and using the broad-band EMIR receivers and the FTS spectrometers in its 200 kHz resolution mode, corresponding to velocity resolutions of 0.17--0.75 \kms. 
The Half Power Beam Width (HPBW) varies between 7\arcsec\ at 350 GHz and 31\arcsec\ at 80 GHz.
%while the main-beam efficiency, $\eta_{\rm mb}$, decreases from 0.82 at 80 GHz to 0.33 in the sub-mm range. 
The data were processed using GILDAS/CLASS90\footnote{http://www.iram.fr/IRAMFR/GILDAS} software.
The spectra were smoothed by a factor of two, baseline subtracted and further resampled at lower spectral resolution if needed to increase the signal-to-noise.
The rms achieved is typically 1.5~mK per interval of 0.4~MHz (i.e. $\sim 1.4$ \kms) across the 3~mm band, and $\sim 5$~mK per interval of $\sim 1$~\kms\ in the 2~mm and 1.3~mm band.
%All the \hcop\ spectra were resampled to a velocity resolution of 1.5 \kms\ to increase the signal-to-noise of the HIFI spectra.
Line intensities are expressed in units of main-beam temperature, T$_{\rm mb}$. We adopted the main-beam efficiencies,  $\eta_{\rm mb}$, monitored by IRAM (http://www.iram.fr). These decrease from 0.82 at 80 GHz to 0.33 in the sub-mm range.
%were converted from antenna temperature, T$_{\rm a}$, to main-beam brightness temperature, T$_{\rm mb}$.

\subsection{{\it Herschel}/HIFI observations}

The IRAM-30m survey is complemented by observations in the sub-millimeter range taken with the Heterodyne Instrument for the Far Infrared (HIFI, \citealt{degraauw10}) on board the {\it Herschel} Space Observatory\footnote{{\it Herschel} is an ESA space observatory with science instruments provided by European-led Principal Investigator consortia and with important participation from NASA.} \citep{pilbratt10} as part of the Guaranteed Time Key Project CHESS \citep{ceccarelli10}. The observations were carried out during 2010 towards the nominal position of B1 and in dual beam switch spectral scanning mode to cover most of the submillimeter bands 1 and 2 (488--628 GHz, 642--794 GHz). Both polarizations (H and V) were observed simultaneously. The receiver was tuned in double sideband (DSB) and the Wide Band Spectrometer (WBS) was used, providing a spectral resolution of 1.1 MHz, corresponding to 0.5--0.7 \kms, which was subsequently degraded to increase the sensitivity. 
The data were processed with the ESA-supported package Herschel Interactive Processing Environment\footnote{HIPE is a joint development by the Herschel Science Ground Segment Consortium, consisting of ESA, the NASA Herschel Sci- ence Center, and the HIFI, PACS and SPIRE consortia.} (HIPE, Ott 2010) version 6 package. Then level 2 fits files were transformed into GILDAS format for baseline subtraction and subsequent sideband deconvolution. The relative calibration between both receivers (H and V) was found very good, and the signals were co-added to improve the signal-to-noise. 
Line intensities were converted from antenna temperature, T$_{\rm a}$, to main-beam brightness temperature, T$_{\rm mb}$, using the main-beam efficiency, $\eta_{\rm mb}$, as determined by \citet{roelfsema12}.

\section{Previous work on L1157}
\label{sect:l1157}

   \begin{figure}
     \centering
     \includegraphics[width=6.cm]{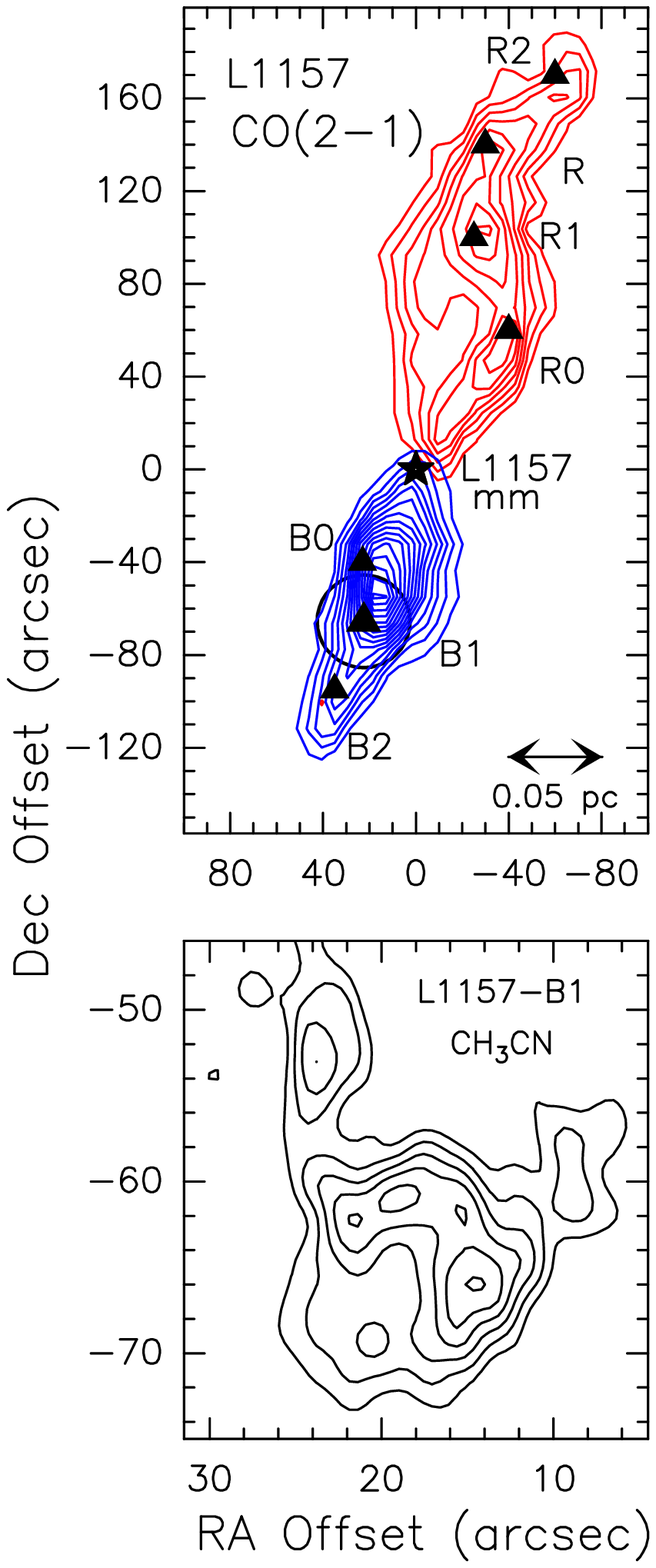}
   \caption{{\it Top panel}: Integrated CO J=2--1 emission of the L1157 bipolar outflow \citep{bachiller01}. Offsets are with respect to the driving source L1157-mm (black star), at coordinates: $\alpha_{\rm J2000}$ = 20$^{\rm h}$ 39$^{\rm m}$ 06$\fs$2, $\delta_{\rm J2000}$ = +68$\degr$ 02$\arcmin$ 16$\farcs$0.
The black triangles and labels indicate the main blue- and red-shifted knots as defined by \citet{bachiller01}. 
Circle is for the largest HPBW of the present dataset (40$\arcsec$), centred on the L1157-B1 bow-shock.
{\it Bottom panel}: The L1157-B1 bow-shock as traced by the CH$_3$CN J=8--7 K=0,1,2 emission at 3~mm, observed with the IRAM PdB interferometer \citep{codella09}.
}
   \label{fig:l1157}
    \end{figure}

The L1157 outflow, located at a distance of $\sim 250$ pc \citep{looney07} and driven by the low mass Class 0 protostar L1157-mm ($\sim 4$ \lsol), is the most chemically rich outflow known so far \citep[e.g., ][]{bachiller97,bachiller01}.
Observations in CO, \hh, and \ho\ lines reveal a highly collimated and precessing bipolar outflow  \citep{davis95,gueth96}, associated with several blue-shifted (B0, B1, B2) and red-shifted (R0, R1, R, R2) bow-shocks \citep{bachiller01,nisini10} (see top panel of Fig.~\ref{fig:l1157}).
Spectral surveys also revealed emission in molecules which are released from dust mantles, such as SO, H$_2$S, OCS, H$_2$CO, CH$_3$OH, \ho, and NH$_3$ \citep{tafalla95,bachiller97,benedettini07,codella10,vasta12}, or sputtered from their core, such as SiO \citep{gueth98,nisini07}.

The brightest bow-shock B1 is an ideal laboratory to investigate the effect of shocks on gas chemistry because is located at $\sim 69$\arcsec\ from the protostar ($\sim$0.1 pc, see bottom panel of Fig.~\ref{fig:l1157}), hence UV irradiation can be neglected.
%observations are not contaminated by emission from the protostar.
The first results obtained from our IRAM-30m/{\it Herschel} survey further highlighted the chemical richness of B1 (\citealt{codella10,lefloch10,benedettini12,codella12a,codella12b}, Busquet et al. submitted). %, G\'omez-Ruiz in preparation).
High-angular resolution interferometric observations reveal that B1 has a clumpy structure with multiple components \citep{benedettini07,codella09,benedettini13}, which cannot be resolved by our single-dish IRAM-30m and {\it Herschel}/HIFI observations.
However, \citet{lefloch12} showed that the analysis of the line profiles provides a simple tool to identify the different spatial and velocity components.
In particular, the CO line profiles from J=1--0 to J=16--15  are well fit by a linear combination of three velocity components ($g_1$, $g_2$, and $g_3$), whose intensity-velocity relation follows an exponential law I(V)~=~I(0)~exp(-|V/V$_0$|), with V$_0$ = 12.5, 4.4, and 2.5~\kms\, for $g_1$, $g_2$ and $g_3$, respectively.
These velocity components trace different regions in the B1 bow-shock, characterized by different sizes and physical conditions:\\
(i) $g_1$: the jet impact region against the cavity ($\sim 7 - 10\arcsec$) in the L1157-B1 bow-shock, with gas at T$_{\rm kin} \sim 210$ K, reaching velocities up to $-40$ \kms; \\
(ii) $g_2$: the outflow cavity associated with B1, made of gas at T$_{\rm kin} \sim 60-80$~K, reaching velocities up to $-20$ \kms;\\
(iii) $g_3$: the cavity associated with the older outflow shock L1157-B2, made of colder gas (T$_{\rm kin} \sim 20$ K) at slower velocities ($\le -10$ \kms).\\
Besides the three components evidenced by the analysis of the CO line profiles, a forth hot and tenuous component (T$_{\rm kin} \sim 1000$ K, \denshh~$\sim 10^4$ \cmc) is detected in \ho\ lines (Busquet et al., submitted).

Interestingly, L1157-B1 also shows emission in molecular ions, such as \hcop, \hcsp, and \nnhp\ \citep{bachiller97,codella10,yamaguchi12,codella13} making it a unique target to investigate the chemistry of molecular ions in protostellar shocks.

\section{Results}
\label{sect:results}

\subsection{Line identification}
%All lines were identified down to the $3\sigma$ level ($\sim 5$~mK in the 3 mm band). 
The ions identified down to the  $3\sigma$ level ($\sim 5$~mK in the 3 mm band) in L1157-B1 are: HCO$^{+}$, H$^{13}$CO$^{+}$, N$_2$H$^{+}$, HOCO$^{+}$,  SO$^{+}$, and HCS$^{+}$.
%One sentence about ions previously detected, and the first detections...
%The detection of HCO$^{+}$ and HCS$^{+}$ in L1157-B1 was first reported by \citet{bachiller97}, while N$_2$H$^{+}$ is reported by \citet{yamaguchi12} and discussed in detail by \citet{codella13}.
Thanks to the high-sensitivity of our survey, several transitions of each molecular species are detected, allowing high accuracy in the derivation of the physical conditions.  
HOCO$^{+}$ and SO$^{+}$ are observed for the first time in L1157-B1 and, more in general, in a protostellar shock.

\begin{figure*}
     \includegraphics[width=\textwidth]{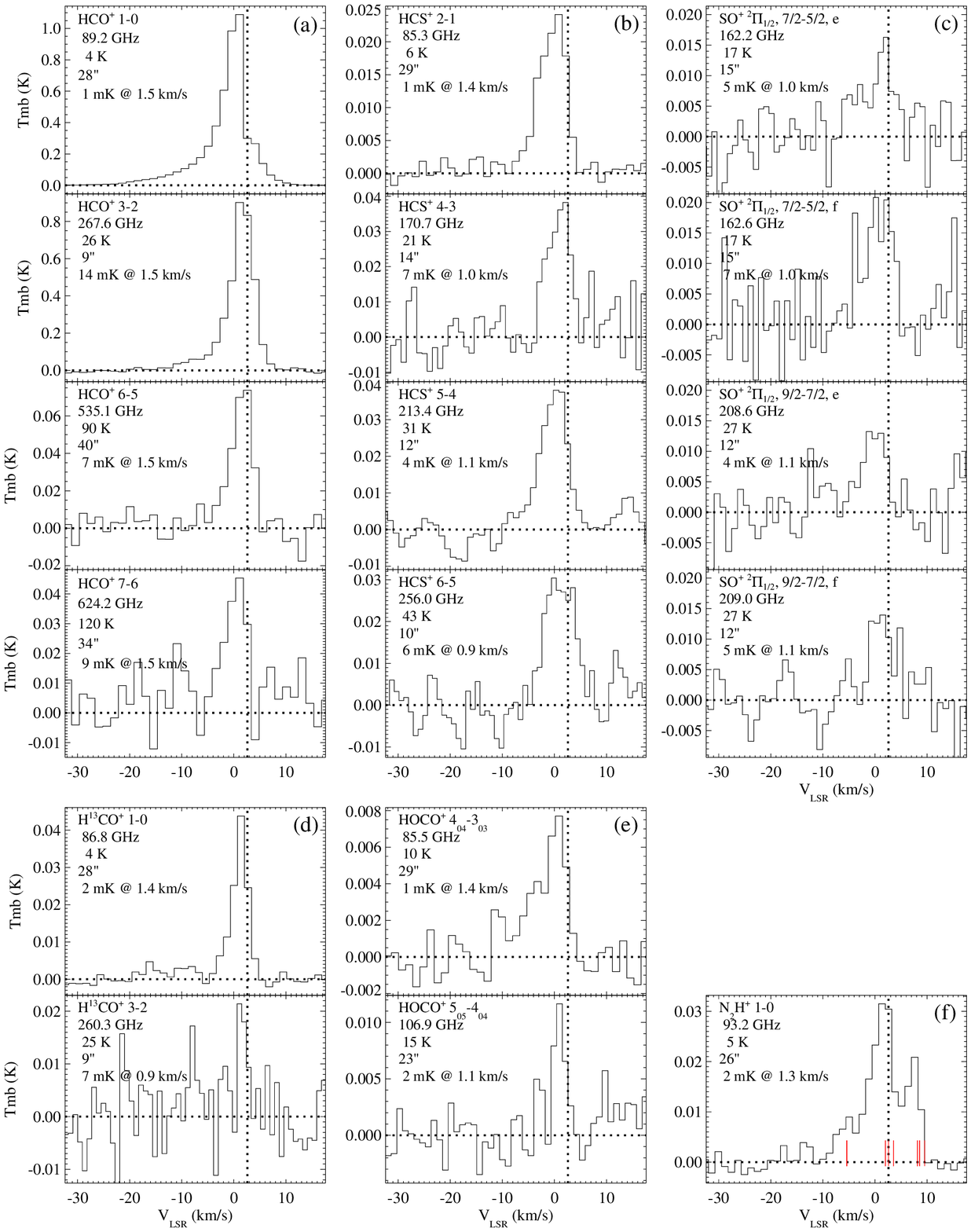}
   \caption{Line profiles of the molecular ions observed in the L1157-B1 shock: \hcop\ (a), \hcsp\ (b), \sop\ (c), \hccop\ (d), \hocop\ (e), \nnhp\ (f). The line intensity is in mean beam temperature (T$_{\rm mb}$). The transition, frequency in GHz, upper level energy in K, HPBW in arcsec, and 1$\sigma$ rms noise in mK are indicated. The baseline and the systemic velocity (V$_{\rm LSR} = + 2.6$ \kms) are indicated by the horizonthal and vertical dotted lines.
For \nnhp\ (f) the 15 hyperfine components of the \nnhp\ 1--0 pattern are indicated by the vertical red lines (due to degeneracy only 7 of the 15 hyperfine lines do not overlap in frequency; of these, only the three main groups are resolved because of blending caused by the line broadening).
%(several of them spectrally unresolved at the present frequency resolution).
%{\bf A FEW BASELINES SHOULD BE CORRECTED.}
%The line profiles are fit by a linear combination of three exponential functions g1 $\propto$ exp(-|V/12.5|) (blue), g2 $\propto$ exp(-|V/4.4|) (red), and g3 $\propto$ exp(-|V/2.0)| (black). The (g1+g2+g3) line fit is overplotted (magenta line).
}
   \label{fig:ions}
    \end{figure*}

The properties of the detected lines (transition, frequency, upper level energy), and their observational parameters (HPBW, rms noise, peak velocity and temperature, Full Width at Half Maximum (FWHM), and integrated intensity) are summarized in Table \ref{tab:lines}. 
The line spectra are shown in Figure \ref{fig:ions}.
All the detected lines peak at blueshifted velocity, $\sim 0.5-3$~\kms\ with respect to systemic, and have a line width of $\sim 3-7$~\kms\ which is consistent with emission originating in the outflow cavities B1 and B2. A detailed analysis of the observed line profiles for each molecular ion is presented in the following sections.

\begin{table*}
\caption[]{\label{tab:lines} Properties of the detected lines from molecular ions: species, transition, frequency ($\nu_{\rm 0}$, in MHz), upper level energy (E$_{\rm up}$, in K), telescope Half Power Beam Width (HPBW, in arcseconds), rms noise (in mK), peak velocity (V$_{\rm peak}$, in \kms) and temperature (T$_{\rm peak}$, in main-beam temperature units), full width at half maximum (FWHM, in \kms) and integrated intensity ($\int {\rm T_{mb} dV}$, in K~\kms). All transitions are observed with IRAM-30m/FTS, except those indicated by the star which are observed with {\it Herschel}/HIFI. A gaussian fit is applied to estimate the line properties for most of the transitions. When the line profile is non-gaussian the line intensity is obtained by integrating the area below the profile. For each species we report a 3$\sigma$ upper limit for the first non-detected transition covered by our observations.\\
%{\bf NO FWHM IS REPORTED WHEN I CANNOT PERFORM A GAUSSIAN FIT. \\
%REPORT THE UPPER LIMIT FOR \nnhp\ 3-2 INSTEAD THAN 6-5.}
}
    \begin{tabular}[h]{cccccccccc}
\hline
Species & Transition & $\nu_{\rm 0}$$^{a}$ & E$_{\rm up}$ & HPBW & rms & V$_{\rm peak}$ & FWHM & T$_{\rm peak}$ & $\int {\rm T_{mb} dV}$ \\
 & & MHz & K & \arcsec & mK & \kms & \kms & mK & K~\kms \\
\hline
\hline
%---------------------------------------------------------------
\hcop
 & J = 1--0$^{\rm ng}$ &  89~188.52 &    4 &   28 &    1 &   1.1 $\pm$ 1.5 &  4.4 $\pm$ 1.5  & 1095 $\pm$    1 & 7.40 $\pm$ 0.01 \\ 
 & J = 3--2$^{\rm ng}$ & 267~557.53 &   26 &    9 &   14 &   1.2 $\pm$ 1.5 &  6.0 $\pm$ 1.5  &  901 $\pm$   14 & 5.49 $\pm$ 0.09 \\ 
 & J = 6--5$^{*~~}$ & 535~061.38 &   90 &   40 &    7 &   1.6 $\pm$ 0.2 &   4.6 $\pm$ 0.4 &   76 $\pm$    7 & 0.38 $\pm$ 0.03 \\ 
 & J = 7--6$^{*~~}$ & 624~208.19 &  120 &   34 &    9 &   0.5 $\pm$ 0.4 &   3.5 $\pm$ 0.8 &   37 $\pm$    9 & 0.14 $\pm$ 0.03 \\ 
 & J = 8--7$^{*~~}$ & 713~342.12 &  154 &   30 &   35 &  -  &  -  &  -  &  $\le$ 0.29 \\ 
%---------------------------------------------------------------
\hccop
 & J = 1--0 &  86~754.29 &    4 &   28 &    2 &   1.2 $\pm$ 0.1 &   3.7 $\pm$ 0.2 &   39 $\pm$    2 & 0.15 $\pm$ 0.00 \\ 
 & J = 3--2 & 260~255.34 &   25 &    9 &    7 &   1.6 $\pm$ 0.3 &   1.8 $\pm$ 0.7 &   23 $\pm$    7 & 0.04 $\pm$ 0.01 \\ 
 & J = 4--3 & 346~998.34 &   42 &    7 &   29 &  -  &  -  &  -  &  $\le$ 0.16 \\ 
%---------------------------------------------------------------
%\hcoop
% & 1--0 &  85~162.16 &    4 &   29 &    1 &  -  &  -  &  -  &  $\le$ 0.009 \\ 
% & 3--2 & 255~480.20 &   25 &   10 &    2 &  -  &  -  &  -  &  $\le$ 0.013 \\ 
%---------------------------------------------------------------
\nnhp$^{b}$
 & J = 1--0         &   93~173.76 &    5 &   26 &    2 & 1.3 $\pm$ 0.1 &  4.3 $\pm$ 0.2  &   29 $\pm$    2 & 0.33$\pm$0.01  \\ 
 & J = 3--2         & 279~511.83 &  27 &     9 &    5 &  -  &  -  &  -  &  $\le$ 0.03  \\ 
% & 6--5$^{*}$ & 558~966.69 &   94 &   38 &   17 &  -  &  -  &  -  &  $\le$ 0.06  \\ 
%---------------------------------------------------------------
\hocop$^{c}$
 & J$_{K_{-1} K_{1}}$ = 4$_{04}$--3$_{03}$$^{\rm ng}$ &  85~531.51 &   10 &   29 &    1 &   0.9 $\pm$ 1.4 &  6.8 $\pm$ 1.4  &    8 $\pm$    1 & 0.06 $\pm$ 0.01 \\ 
 & J$_{K_{-1} K_{1}}$ = 5$_{05}$--4$_{04}$$^{\rm ng}$ & 106~913.57 &   15 &   23 &    2 &   1.0 $\pm$ 1.1 &  3.3 $\pm$ 1.1  &   12 $\pm$    2 & 0.04 $\pm$ 0.01 \\ 
 & J$_{K_{-1} K_{1}}$ = 6$_{06}$--5$_{05}$$^{~~~}$ & 128~295.06 &   22 &   19 &   11 &  -  &  -  &  -  &  $\le$ 0.07 \\ 
% & 7$_{07}$ -- 6$_{06}$ & 149~675.88 &   29 &   16 &    3 &  -  &  -  &  -  &  $\le$ 0.03 \\  %0.6 $\pm$ 1.0 &   9.5 $\pm$ 2.0 &    7 $\pm$    3 & 0.07 $\pm$ 0.01 \\
%---------------------------------------------------------------
%\sop\ $^2 \Pi_{1/2}$
\sop$^{d}$
 & $^2\Pi_{1/2}$, J = ~~5/2--3/2, l = e$^{~~~}$ & 115~804.41 &    9 &   21 &   13 &  -  &  -  &  -  &  $\le$ 0.09  \\ 
 & $^2\Pi_{1/2}$, J = ~~5/2--3/2, l = f$^{~~~}$ & 116~179.95 &    9 &   21 &   15 &  -  &  -  &  -  &  $\le$ 0.12  \\ 
 & $^2\Pi_{1/2}$, J = ~~7/2--5/2, l = e$^{\rm ng}$ & 162~198.59 &   17 &   15 &    5 &   2.0 $\pm$ 1.0 &  2.9 $\pm$ 1.0  &   13 $\pm$    5 & 0.09 $\pm$ 0.02 \\ 
 & $^2\Pi_{1/2}$, J = ~~7/2--5/2, l = f$^{~~~}$ & 162~574.06 &   17 &   15 &    7 &   0.6 $\pm$ 0.8 &   6.3 $\pm$ 1.5 &   17 $\pm$    7 & 0.12 $\pm$ 0.03 \\ 
 & $^2\Pi_{1/2}$, J = ~~9/2--7/2, l = e$^{~~~}$ & 208~590.03 &   27 &   12 &    4 &  -0.2 $\pm$ 0.6 &   5.7 $\pm$ 1.7 &   13 $\pm$    4 & 0.08 $\pm$ 0.02 \\ 
 & $^2\Pi_{1/2}$, J = ~~9/2--7/2, l = f$^{~~~}$ & 208~965.42 &   27 &   12 &    4 &   1.3 $\pm$ 0.7 &   5.6 $\pm$ 2.0 &   13 $\pm$    4 & 0.08 $\pm$ 0.02 \\ 
 & $^2\Pi_{1/2}$, J = 11/2--9/2, l = e$^{~~~}$ & 254~977.94 &   39 &   10 &    5 &  -  &  -  &  -  &  $\le$ 0.03 \\ 
 & $^2\Pi_{1/2}$, J = 11/2--9/2, l = f$^{~~~}$ & 255~353.23 &   39 &   10 &    5 &  -  &  -  &  -  &  $\le$ 0.03 \\ 
% & 13/2--11/2, e & 301~361.50 &   53 &    8 &    9 &  -  &  -  &  -  &  $\le$ 0.09  \\ 
% & 13/2--11/2, f &	301~736.79 &   54 &    8 &   11 &  -  &  -  &  -  &  $\le$ 0.09  \\ 
%---------------------------------------------------------------
\hcsp
 &  J = 2--1 &  85~347.90 &    6 &   29 &    1 &  -0.3 $\pm$ 0.8 &   6.2 $\pm$ 1.9 &   23 $\pm$    1 & 0.15 $\pm$ 0.04 \\ 
 &  J = 4--3 & 170~691.73 &   21 &   14 &    7 &   0.9 $\pm$ 0.3 &   5.6 $\pm$ 0.7 &   36 $\pm$    7 & 0.21 $\pm$ 0.02 \\ 
 &  J = 5--4 & 213~360.53 &   31 &   12 &    4 &  0.0 $\pm$ 0.2 &   6.2 $\pm$ 0.5 &   38 $\pm$    4 & 0.25 $\pm$ 0.02 \\ 
 &  J = 6--5 & 256~027.80 &   43 &   10 &    6 &   1.1 $\pm$ 0.3 &   6.9 $\pm$ 0.7 &   30 $\pm$    6 & 0.22 $\pm$ 0.02 \\ 
 &  J = 7--6 & 298~690.88 &   57 &    8 &   16 &  -  &  -  &  -  &  $\le$ 0.09 \\ 
%---------------------------------------------------------------
%\cop
% & J=1-0, F=1/2-1/2 & 117.7 &    6 &   21 &   95 &  -  &  -  &  -  &  -  \\ 
% & J=2-1, F=3/2-3/2 & 235.4 &   17 &   10 &    3 &  -  &  -  &  -  &  -  \\ 
% & J=2-1, F=3/2-1/2 & 235.8 &   17 &   10 &    5 &  -  &  -  &  -  &  -  \\ 
% & J=2-1, F=5/2-3/2 & 236.1 &   17 &   10 &    4 &  -  &  -  &  -  &  -  \\ 
%---------------------------------------------------------------
\hline \\
    \end{tabular}\\
\small
$^{*}$observed with {\it Herschel/}HIFI.\\
$^{\rm ng}$non-gaussian profile. The line intensity is obtained by integrating the area below the profile. \\
$^{a}$frequencies are from the JPL molecular database \citep{pickett98}, and from the CDMS database \citep{muller01} for \nnhp. \\
$^{b}$the line properties quoted for the \nnhp\ 1--0 and 3--2 transitions refer to the brightest of their 15 and 45 hyperfine components, i.e. the F1, F = 2,3--1,2 and the F1, F = 4,5--3,4 components, respectively. The line intensity is integrated over all the components.\\  
%the line properties quoted for the \nnhp\ 6--5 transition refer to the brightest of its 38 hyperfine components, i.e. the F1, F = 7,8--6,7 components
$^{c}$the transitions of \hocop\ are characterized by the quantum numbers J, which is the total angular momentum, and $K_{-1}, K_{1}$, which are the angular momenta around the molecule's symmetry axis.\\ 
$^{d}$\sop\ is a reactive radical with $^2\Pi_{1/2}$ ground state, whose spin-doubled rotational transitions, characterized by the quantum number J, are further split by $\Lambda$-type doubling.\\ 

\end{table*}

\subsection{Physical conditions and abundances}
\label{sect:phys}

The gas physical conditions and the column density and abundance of each molecular species were derived from a multi transition analysis, by means of  a radiative transfer code in the LVG approximation whenever more than one line is detected and the collisional coefficients are available (\hcop, \hocop, and \hcsp), and in the hypothesis of Local Termodinamic Equilibrium (LTE) otherwise (\nnhp, and \sop). 
% FROM BERTRAND: As for HCS$^{+}$, the only collisional coefficients published in the literature (Monteiro et al, 1984) cover only a limited range of temperature (5 to 60~K) and strongly lack of accuracy. In practice, they can be reasonably well approximated by the CS collisional coefficients (Turner et al. 1992) enhanced by a factor of 1.5.
% FROM CECILIA: (Lacking a more adequate coverage of temperatures, we adopted the collisional coefficients of the CS-He. In LVG_GRE we have the coefficients by Lique et al. 2006.) Lacking recent computations on the HCS+ collisional coefficients, we followed Turner et al. (1992) and used instead the CS-H2 coefficients, that are computed by Lique et al. (2006) and that we retrieved from the BASECOL database (give the web site: Dubernet et al. 2012).
The Einstein coefficients and upper level energies are retrieved from the Jet Propulsion Laboratory (JPL) database\footnote{http://spec.jpl.nasa.gov}  \citep{pickett98} and the Cologne Database of Molecular Spectroscopy (CDMS)\footnote{http://www.astro.uni-koeln.de/cdms} \citep{muller01}, while the collisional coefficients are taken from the BASECOL database\footnote{http://basecol.obspm.fr/} \citep{dubernet13}. 
In particular, the reference for the collisional coefficients are: 
\citet{flower99} for \hcop;
\citet{hammami07} for \hocop;
\citet{monteiro84} for \hcsp.
In the case in which an LTE analysis is applied the main source of uncertainty on the derived abundance is the assumed excitation temperature and size of the emitting region. This leads to an uncertainty of up to an order of magnitude for \nnhp, and about a factor three for \sop, for which the excitation temperature is derived by fitting the detected transitions.
When, instead, an LVG analysis is applied the uncertainty on the abundance is obtained by considering the difference between the best-fit and the second best-fit solutions and is around a factor three.
The gas kinetic temperature and \hh\ density, T$_{\rm kin}$ and \denshh, the column density, N$_{\rm species}$, and fractional abundances with respect to \hh, X$_{\rm species}$, derived for the analysed molecular ions are summarized in Tab.~\ref{tab:lvg}.

\subsubsection{\hcop\ and isotopologues}

%   \begin{figure}
%     \includegraphics[width=8.cm]{tau.ps}
%   \caption{From top to bottom panels the figure show the \hcop\ and \hccop\ 1--0 profiles, their intensity ratio, and the \hcop\ 1--0 line optical depth derived assuming an isotopic ratio of 77 \citep{wilson94}.}
%   \label{fig:tau}
%    \end{figure}

Figs.~\ref{fig:ions}a and \ref{fig:ions}d show all the detected transition from \hcop\ and its isotopologue \hccop.
\hcop\ has been widely detected in prestellar cores \citep[e.g., ][]{caselli98}, in embedded low-mass protostars, and along their outflows \citep[e.g., ][]{hogerheijde98}.
\hcop\ emission is detected up to the J = 7--6 transition. 
%Thanks to the high sensitivity of the present IRAM-30m observations, 
We also detect emission from \hccop\ in the 1--0 and, tentatively, in the 3--2 line (the latter with a signal-to-noise of 4). 
The \hcoop\, 1--0, and 3--2 lines, instead, are not detected. 
The \hcop\ 1--0 line shows bright and very broad wings extending up to $-40$ \kms.
%, i.e. up to velocities much higher than previously reported by BP97 (V$_{max} \sim -10$ \kms).
So far, only CO, H$_2$O, and SiO have been observed at these extreme velocities in L1157-B1 \citep{lefloch10,lefloch12}.
High-velocity emission in \hcop\ lines has been detected also in the outflows driven by the Class 0 protostars L1448-mm and IRAS 04166+2706 \citep{tafalla10}.

%Figure \ref{fig:tau} show the \hcop\ and \hccop\ 1--0 profiles, their intensity ratio, and 
The \hcop\ 1--0 line optical depth is derived from the \hcop\ to the \hccop\ 1-0 line ratio assuming an isotopic ratio of 77 \citep{wilson94}.  
%The line of the main isotopomer exhibits an absorption dip, which is due to contamination from emission at the reference position, located at 3 arcmin from the protostar. 
A simple LTE analysis of the $^{12}$C/$^{13}$C line ratio, R, indicates that the \hcop\ 1--0 emission is optically thick at the cloud velocity (R~$\sim 11$, $\tau \sim 7$) and becomes optically thin in the wings (R~$\sim 68$, $\tau \sim 0.3$ at $\sim -2$ \kms). 

%Previous analysis of CO emission in L1157-B1 indicates that all CO line profiles from J=1--0 to J=16--15 are well fit by a linear combination of three components (g1, g2, and g3) which follows an exponential law $\propto$ exp( $-| V / V_0 |$ ), with V$_0 =$ 12.5, 4.4, and 2.5 \kms\ \citep{lefloch12}.
%The three exponentials are believed to trace different shocked gas components which have different sizes and temperatures: 
%i) a small region ($\sim$10\arcsec) where the primary jet impacts on the L1157-B1 bow-shock (T$\sim 210$ K), 
%(ii) the outflow cavity associated with B1 ($\sim$20\arcsec, T$\sim 64$ K), and 
%(iii) the older cavity L1157-B2 (T$\sim 23$ K).
%\citet{lefloch12} finds that the warm gas in the outflow cavity, g2, is producing the bulk of the emission for all CO lines. However, an hotter component associated with the primary jet, g1, is needed to reproduce the line emission at high velocities, while emission from the older cavity B2 is important close to systemic velocity.

\begin{figure}
     \includegraphics[width=8.8cm]{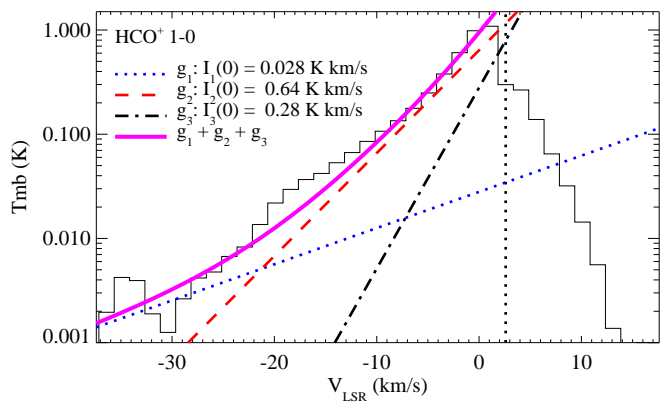}
   \caption{Profile of the \hcop\ 1--0 line observed in the L1157-B1 shock. The line intensity is in mean beam temperature (T$_{\rm mb}$) and the systemic velocity (V$_{\rm LSR} = +2.6$ \kms) is indicated by the vertical dotted lines. The line profile is fit by a linear combination of three exponential functions $g_1$ = I$_1$(0) exp(-|V/12.5 km/s|) (blue dotted line), $g_2$ = I$_2$(0) exp(-|V/4.4 km/s|) (red dashed line), and $g_3$ = I$_3$(0) exp(-|V/2.5 km/s)| (dotted-dashed black line). The ($g_1$ + $g_2$ + $g_3$) line fit is overplotted (magenta solid line).
}
   \label{fig:hcop_components}
    \end{figure}

We tentatively fit the profile of the \hcop\ 1--0 line using the three exponential components as for CO.
A good fit is obtained adopting for the three components the zero velocity intensities summarized in Fig.~\ref{fig:hcop_components}.
Similarly to what has been found for the CO lines \citep{lefloch12}, the bulk of the emission originate in the outflow cavity B1 ($g_2$), while the $g_1$ jet component is dominant at high velocities (V $< -20$ \kms), and the $g_3$ component, accounting for the emission from the old cavity B2, is contributing only close to systemic velocity.
The  emission excess detected between $-20$ \kms\ and $-15$ \kms\ could be related to the compact clumps along the cavity detected with high-resolution interferometric observations and showing a second emission peak at those velocities \citep{benedettini13}. 
The higher-J \hcop\ lines are detected with lower signal-to-noise, thus the $g_1$ component remains undetected and they are fit by a combination of the $g_2$ and $g_3$ components.

We use an LVG model to fit the fluxes of all the \hcop\ lines, after subtracting the $g_1$ component from the \hcop\ 1--0 line.
The flux of the different lines is well accounted for by extended emission from a single gas layer at T$_{\rm kin} \sim 60$ K, \denshh~$\sim 10^5$ \cmc, and a column density \nhcop~$\sim 6 \ 10^{12}$ \cms. This solution is in very good agreement with the size and physical conditions found for the outflow cavity B1 ($g_2$ component) for the CO lines \citep{lefloch12} (see Table \ref{tab:lvg}). Adopting the CO column density found by these authors, N$_{\rm CO} = 9 \ 10^{16}$ \cms, we estimate an \hcop\ relative abundance \xhcop($g_2$)~= \nhcop/\nhh $\sim 7 \ 10^{-9}$.

A tentative estimate of the \hcop\ abundance in the jet component $g_1$, can be obtained by assuming the same physical conditions as for CO (T$_{\rm kin} =210$ K, \denshh $\ge 10^{6}$ \cmc) and computing the \hcop\ column density required to account for the $g_1$ integrated area. Adopting $\Delta \rm V =10$ \kms, we estimate a column density \nhcop~$\sim 1-3 \ 10^{12}$ \cms, depending on the assumed density (\denshh $ = 10^{6} - 10^{7}$, respectively), implying an \hcop\ abundance \xhcop($g_1$)~$\sim 1 - 3 \ 10^{-8}$. 
The abundance estimated in the $g_1$ component is in agreement within a factor of two with the estimate by \citet{bachiller97}.  
We recall that the latter is derived using only the 1--0 line and assuming that the emission is optically thin and thermalized (T$_{\rm kin} = 80$ K).
  
Within the uncertainties of the present and previous studies by \citet{bachiller97}, the \hcop\ abundance estimated both in the jet impact region ($g_1$) and in the outflow cavities ($g_2$) can be considered in agreement with  what estimated at the source position by \citet{bachiller97} within a factor of a few.
This is in agreement with previous observations of outflows in the \hcop\ lines showing no evident enhancement of the \hcop\ abundance both in the outflow cavities and in the high velocity component with respect to dense core value \citep{hogerheijde98,tafalla10}. 
The chemistry of \hcop\ and its evolution in the shock will be discussed in Sect.~\ref{sect:astrochem}. 

\subsubsection{\nnhp}

A detailed analysis of the detected \nnhp\ 1--0 line (see Fig.~\ref{fig:ions}f) is presented in \citet{codella13}.
These authors show that the observed line profile is consistent with emission by either the B1 outflow cavity ($g_2$) and/or with the older and colder B2 cavity ($g_3$).
The peak velocity and line width (V$_{\rm peak} \sim + 1.3$~\kms, FWHM~$\sim 4.3$~\kms) are in agreement with those found for the other molecular ions.
The colum density and fractional abundance (\nnnhp~$\sim 0.4-8 \ 10^{12}$~\cms, \xnnhp~$\sim 0.4-8 \ 10^{-9}$) are derived by assuming LTE conditions at a temperature of $\sim 20-70$~K and an emitting size of $\sim 20-25$\arcsec\ as derived from the analysis of the CO lines by \citet{lefloch12}.
As for \hcop, also \nnhp\ is not enhanced in the shock with respect to the on-source value estimated by \citet{bachiller97}. 
The chemistry of \nnhp\ is further discussed in Sect.~\ref{sect:astrochem}.

\subsubsection{\hocop}
\label{sect:hocop}

\begin{table*}
%\small
\caption[]{\label{tab:lvg} Gas physical conditions (kinetic temperature, T$_{\rm kin}$, and \hh\ density, \denshh), column density, N$_{\rm species}$, and fractional abundances, X = N$_{\rm species}$/\nhh, for CO \citep{lefloch12} and the observed molecular ions as derived from an LVG (for \hcop, \hocop, and \hcsp) or LTE (for \nnhp\ and \sop) analysis of the detected lines. The values derived for \nnhp\ are from \citet{codella13}.}
    \begin{tabular}[h]{ccccccc}
\hline
Species & Cmp           & T$_{\rm kin}$ & \denshh & N$_{\rm species}$ & X = N$_{\rm species}$/\nhh \\% & f=X/X$_{s}$ \\ %& X$_{cloud}$           & X$_{cloud}$           &X$_{shock}$ \\
           &                     & K & \cmc      & \cms &                            \\% &                \\ %      & \citet{herbst86} & Flower \& Pineau &              \\
%           &                     &  &       &  &                            &                      &   $10^4$ \cmc, $10-50$ K & $10^2-10^6$ \cmc, $10$ K &              \\

\hline
\hline
CO$^{a}$ & $g_1$ & 210       & $\ge$ 10$^{6}$   & 9  10$^{15}$ &  10$^{-4}$  \\% & \\ %& &\\
              & $g_2$ & 60--80 & $\ge$ 10$^{5}$     & 9  10$^{16}$ &  10$^{-4}$ \\% & \\ %& & \\ 
              & $g_3$ & 20         & $\ge$ 10$^{5}$    &  1 10$^{17}$ &   10$^{-4}$ \\% & \\ %& & \\
\hline
\hcop & $g_1$ & 210$^{b}$  & 10$^{6}$-10$^{7}$$^{b}$ & 1--3 10$^{12}$ & 1--3 10$^{-8}$  \\% & 9 \\ %& $2 \ 10^{-10}$(HM) $- 2.4 \ 10^{-8}$(LM) &  $2 \ 10^{-9} - 2.4 \ 10^{-8}$   &    10$^{-8}$$^{d}$ \\
           & $g_2$ & 60    & 10$^{5}$                & 6 10$^{12}$ & 7 10$^{-9}$  \\% & 4.7 \\ %& & \\
\hline
\nnhp$^{c}$  & $g_2$/$g_3$ & 20--70$^{c}$  &   -  & 0.4--8 10$^{12}$ & 0.4--8 10$^{-9}$ \\% & 0.5-2 \\ %& $2 \ 10^{-12} - 8 \ 10^{-10}$(LM) & $3 \ 10^{-11} - 10^{-8}$ & \\ 
\hline
\hocop        &  $g_2$/$g_3$ & 35     & 10$^{4}$             & 1 10$^{12}$  & 1 10$^{-9}$  \\% & \\ %& $2 \ 10^{-14} - 2 \ 10^{-11}$(LM) & & \\
\hline
%\sop$^{d}$   & $g_1$ & 10        &              -            & 2 10$^{12}$ & 2 10$^{-8}$ \\% &  \\ %& $7 \ 10^{-15} - 3 \ 10^{-11}$(LM) & $4 \ 10^{-10} - 10^{-8}$ & 10$^{-9}$--10$^{-8}$$^{e}$\\ 
\sop$^{d}$   & $g_2$/$g_3$ & 25$^{d}$ &              -            & 7 10$^{11}$ & 8 10$^{-10}$ \\% &  \\%& & \\
\hline
\hcsp & $g_2$ & 80       & 8 10$^{5}$      & 6 10$^{11}$ & 7 10$^{-10}$ \\% &  \\%& $4 \ 10^{-13}$(LM) $ - 10^{-10}$(HM) &  $3 \ 10^{-11} - 7 \ 10^{-10}$   & 5.6 10$^{-9}$$^{d}$  \\ 
          & $g_3$ &  20$^{e}$      & 10$^{5}$$^{e}$         & 3 10$^{11}$ & 3 10$^{-10}$ \\% &  \\ %& &\\
\hline
    \end{tabular}\\
\small
$^{a}$ \citet{lefloch12}. The fractional abundance of CO is assumed \\
$^{b}$ T$_{\rm kin}$ and \denshh\ in $g_1$ are assumed to be as derived from CO \\
$^{c}$ \nnnhp, \xnnhp\ estimated by \citet{codella13} assuming LTE at T$_{\rm kin} =20-70$~K and a source size of $\sim 20-25$\arcsec\ as inferred from CO \\
$^{d}$ Since no collisional coefficients are available we apply an LTE analysis assuming that the emission originates in the extended outflow cavities ($g_2, g_3$). Hence, at difference with the other ions, the estimated temperature is an excitation temperature, T$_{\rm ex}$\\ 
%or from the compact jet impact region ($g_1 \sim 7-10$\arcsec). \\
$^{e}$ T$_{\rm kin}$ and \denshh\ in $g_3$ are assumed to be as derived from CO. This assumption relies on the similarity of CO and \hcsp\ profiles at low velocities and on the results of the LVG analysis (see text). \\
%$^{d}$ \citet{viti02}\\
%$^{e}$ \citet{neufeld89a}\\
%\normalsize
\end{table*}

We detect \hocop\ emission in the $4_{04} - 3_{03}$ and $5_{05} - 4_{04}$ lines at 85.5 and 106.9 GHz (see Fig.~\ref{fig:ions}e). 
So far \hocop\ has been detected in different environments, i.e. in the Galactic center region \citep{thaddeus81,minh88,minh91,deguchi06}, in translucent and dark clouds \citep{turner99},  and around the low-mass Class 0 protostar IRAS~04368+2557 in L1527 \citep{sakai08}. 
The reported detection, however, is the first one directly associated with a shock spot.
%This is the first detection of HOCO+ directly associated with a shock spot and could be a strong confirmation of the fact that, as suggested by previous studies, its abundance is enhanced in shocks, thus making HOCO+ an effective shock tracer.
%Assuming that the two lines are excited in the primary jet component g1, where the physical conditions are as derived from CO (i.e. T=210 K), and that they are in thermal equilibrium and optically thin, we derive a column density of $\sim$10$^{13}$ \cms\ and a abundance of X=1 10$^{-7}$.\\

The signal-to-noise of the detected lines is quite low ($\sim 4-6$). 
Similarly to the other ions, the profiles favor an origin in the outflow cavities B1 and/or B2 ($g_2$, $g_3$). 
However, the lower excitation \hocop\ $4_{04} - 3_{03}$ line at 85.5 GHz also shows a blueshifted wing extending up to $-12$ \kms.
%, which could be associated with the jet impact region, $g_1$.}
From an LVG analysis including the two detected lines and the upper limit for the $6_{06} - 5_{05}$ line at 128.3 GHz we find that the emission is extended and the best fit is obtained for \denshh~$\sim 10^4$ \cmc, T$_{\rm kin} \sim 35$ K, \nhocop~$\sim 10^{12}$ \cms.
Thus, if we consider a CO column density of $\sim 10^{17}$ ($g_2$, $g_3$) we find \xhocop~$\sim 10^{-9}$.

The estimated abundance is 1-2 orders of magnitude larger than  what predicted by gas-phase chemistry in cold quiescent clouds \citep{herbst86,turner99}.
Previous studies suggest that \hocop\ may be enhanced in shocks and/or UV irradiated regions due to the release of \coo\ from dust grain mantles, then reacting with \hhhp\ to form \hocop\ \citep{herbst77,minh91,deguchi06,sakai08}.
However, those studies are based on unresolved observations of complex environments and do not allow to conclude on the origin of \hocop.
Our detection in the L1157-B1 shock, located 69\arcsec\ away from source, is a strong confirmation that \hocop\ is an effective shock tracer. 
%and may be a useful probe of interstellar \coo, which lacks a permanent electric dipole moment making radio detection impossible.
The chemistry \hocop\ and the origin of the observed emission is further discussed in Sect.~\ref{sect:astrochem}.

\subsubsection{\sop}

Fig.~\ref{fig:ions}c shows the \sop\ transitions detected by our IRAM-30m observations.
Up to date \sop\ lines have been detected in the shocked molecular clump associated with the supernova remnant IC 443G \citep{turner92}, in cold clouds and warm active star froming regions \citep{turner94}, and in translucent molecular clouds \citep{turner96}.
The reported detection is the first detection of \sop\ emission directly associated with a protostellar shock.
%%The \sop\ doublets at 162.6 GHz (E$_{up} \sim 17$ K) is clearly observed in the 2 mm band, while the emission from the %%lower excitation doublets at 115.8, 116.2 GHz (E$_{up} \sim 9$ K) is not detected possibly due to the high rms noise in %%this region of the 3 mm band (rms $\sim 13-15$ K).
We could not detect the \sop\ doublet at 115.8, 116.2 GHz in the 3mm band (E$_{\rm up} \sim 9$ K) because of the high rms noise (about 13--15~mK). The lines from the higher excitation doublets at 162.2, 162.6 GHz (E$_{\rm up} \sim 17$ K), at 208.6, 209.0 GHz (E$_{\rm up} \sim 27$ K) are detected.
%, and at 255.0, 255.4 GHz (E$_{\rm up} \sim 39$ K) are detected, with the exception of the 255.4 GHz line which could be hidden in the noise.

The line profiles suggest that they are dominated by emission from the outflow cavities ($g_2$, $g_3$) with no evidence of high velocity emission from the jet impact region ($g_1$).
However, the emission is rather weak and the signal to noise ratio is not high enough to firmly conclude about the origin of the emission from a simple line profile analysis.
Since no collisional coefficients are available, we derive the \sop\ abundance from the detected line fluxes under the hypothesis of LTE.
The \sop\ abundance estimated under this hypothesis depends on the size of the emitting region, thus we consider two limiting-cases. 
%either the emission is extended, i.e. from the outflow cavity ($g_2$), or the emission is more compact, as for the $g_1$ component.
%%
%% Here, we first obtain N(SO+), which is estimated in a beam of 11'' - 15''
%% We then divide by N(CO) averaged over the same beam
%% Source-averaged N(CO)= 8e16 cm-2 ; in a beam of 15'' : N(CO) = 5e16 cm-2

If we assume that \sop\ arises from the outflow cavities, either B1 ($g_2$) or B2 ($g_3$), the emission is more extended than the beam size, and, as first order, the main-beam brightness temperature is a reasonable approximation of the intrinsic source brightness temperature.
The relative intensities of the doublets at 162~GHz  and 208~GHz imply an excitation temperature of $\sim 25$~K; we estimate a column density \nsop~$\sim 7 \ 10^{11}$ \cms\ and a relative abundance \xsop~$\sim 8 \ 10^{-10}$.

If instead we assume that \sop\ originates from the compact jet impact region ($g_1 \sim 7-10$\arcsec), then one has to correct for the source coupling with the telescope beam. 
%This effect is more pronounced and varies more strongly with frequency when the emitting region has a small size. 
We estimate the source coupling factor from an SiO J=8--7 map at $7.4\arcsec$ resolution obtained at the IRAM-30m telescope, convolved at the resolution of the different \sop\ transitions. 
%We found that, convolved at the resolution of the 7/2--5/2 line, T$_{\rm mb}$(9/2--7/2) $\sim 11$~mK. Our LTE analysis then yields T$_{\rm ex} \sim 15$~K and a beam-averaged column density \nsop~$\sim 7~10^{11}$ \cms. This corresponds to a source-averaged abundance \xsop~$\sim 2~10^{-8}$. 
Our LTE analysis then yields T$_{\rm ex} \sim 10$~K and a column density \nsop~$\sim 2~10^{12}$ \cms, corresponding to a relative abundance \xsop~$\sim 2~10^{-8}$. 
The derived excitation temperature is much lower than what estimated from the CO lines in the $g_1$ component by \citep{lefloch12} (T$_{\rm ex} \sim 210$ K).
This suggests that, similarly to what found for the other ions, the bulk of the observed \sop\ emission originates from the B1, B2 cavities.

To summarize, we estimate an \sop\ abundance of $\sim 8 \ 10^{-10}$ in the outflow cavities.
This value is 1--2 orders of magnitude larger than what predicted in cold clouds \citep{herbst86} and favor a shock origin, as suggested by previous detections of large \sop\ abundances in active star forming regions \citep{turner94}.
The chemistry of \sop\ is discussed in Sect.~\ref{sect:astrochem}.

\subsubsection{\hcsp}

Fig.~\ref{fig:ions}b shows the \hcsp\ lines detected with the IRAM-30m telescope.
\hcsp\ has been previously detected in molecular clouds \citep[e.g., ][]{ohishi92,sutton95}, in the low-mass protostar IRAS 16293–2422 \citep[e.g., ][]{schoier02a}, and in association with young sources driving outflows in Cepheus A \citep{codella05}. 
Emission in the bow-shock L1157-B1 was already reported by \citet{bachiller97} and \citet{yamaguchi12}. 
We detected the  2--1, 4--3, 5--4, and 6--5 lines with a signal-to-noise $> 4$, while only an upper limit is derived for the 7--6 line.

%%%%%% VERSION 1
%The line profile can be reasonably well reproduced by a relation of the type $I(v)= I(0)\exp(-|v/4.4|)$, similar to the signature of the CO gas from the B1 outflow cavity. However, a $g_3$ component, i.e. an additional contribution by the old B2 cavity, is needed to reproduce the emission close to systemic velocity.
%In particular, the $g_3$ component is largely contributing to the 2--1 line, while it is negligible in the higher-J lines.
%%%%%% VERSION 2
The signal to noise ratio is too low to perform a proper line profile analysis.
No high velocity emission associated with the $g_1$ component is detected and the lines appear to be dominated by emission from the outflow cavity B1 ($g_2$). 
Additional emission by the old B2 cavity ($g_3$) could contribute close to systemic velocity. 
%%%%%% VERSION 3
%{\bf Despite the signal to noise is much lower than for the \hcop\ lines we tentatively fit the \hcsp~2--1 line profile using the three exponential components as for CO (see Fig.~\ref{fig:}, bottom panel).
%No high velocity emission associated with the $g_1$ component is detected, instead the line is well reproduced by emission from the outflow cavity B1 ($g_2$), with the same exponential law as for CO. 
%A $g_3$ component, i.e. an additional contribution by the old B2 cavity, is needed to reproduce the emission close to systemic velocity.
%The zero velocity intensities of the $g_2$ and $g_3$ components are summarized in Fig.~\ref{fig:} }

Our LVG calculations confirm that the emission is dominated by the $g_2$ component. %the interpretation suggested by the line profile fitting.
All the lines but the 2--1 are well reproduced by a single extended component with T$_{\rm kin} \sim 80$ K, \denshh $\sim 8 \ 10^5$ \cmc, and \nhcsp~$\sim 6 \ 10^{11}$ \cms.
The estimated physical conditions are consistent with those found for the B1 outflow cavity (i.e. the $g_2$ component), as derived from CO, and indicate an \hcsp\ abundance of $7 \ 10^{-10}$.
However, the \hcsp\ 2--1 observed line intensity is about twice as large as the one predicted by the LVG model for the $g_2$ component. In order to account for the intensity of the J=2--1, it is necessary to consider an additional component, of low excitation. Based on the \hcsp~2--1 line profile this component could be associated with the B2 cavity,  thus we assume the same physical conditions as derived from CO, i.e. T$_{\rm kin} \sim 20$ K, \denshh~$\sim 10^5$ \cmc. We find that for \nhcsp~$\sim 3 \ 10^{11}$ \cms\ we obtain a good fit of the 2--1 line, while the contribution to the higher-J lines is negligible.

To summarize, we find that the \hcsp\ profiles are well produced by the emission of the B1 and B2 outflow cavities, with similar abundances \xhcsp\ of 7 and $3 \ 10^{-10}$, respectively.
These values are a factor four up to one order of magnitude lower than what estimated by \citet{bachiller97} based on tentative detections of the 2--1, 5--4 lines.
As for the other molecular ions, the chemistry of \hcsp\ is discussed in Sect.~\ref{sect:astrochem}.

%{\bf \hcsp\ abundances are strongly dependent on the assumed S in gas-phase, see Fig.~\ref{fig:astrochem}. The S gas-phase and thus the \hcsp\ abundance can be increased in the shock.}

%Similarly to \hcop\, the estimated \hcsp\ abundance can be produced by gas-phase chemistry in cold dense clouds for n $\sim 10^3- 10^4$ \cmc.
%See pre-shock abundances predicted by Flower and Pineau des Forets for T $=10$ K (see Fig.~\ref{fig:abu_dens}). 
%These are in agreement with abundance previously estimated by \citet{herbst86} (X(\hcsp) up to $10^{-10}$ for n $= 10^4$ \cmc).

%{\bf formation routes} \\
%\hhs\ + \cp\     $\rightarrow$ H      + \hcsp\ \\
%\hhs\ + \ccnp\ $\rightarrow$ HCN + \hcsp\ \\
%\chh\ +  \splus\  $\rightarrow$ H      + \hcsp\ \\
%\hhs\ + \chp\   $\rightarrow$ \hh\ + \hcsp\ \\
  
%{\bf abundance in quiescent clouds:} 4 10$^{-13}$ -- 10$^{-10}$ \citep{herbst86}\\

%{\bf abundance in UV-producing shocks:} 5.6 10$^{-9}$ \citep{viti02}\\

\section{Chemistry of molecular ions}
\label{sect:astrochem}

  \begin{figure*}
     \includegraphics[width=\textwidth]{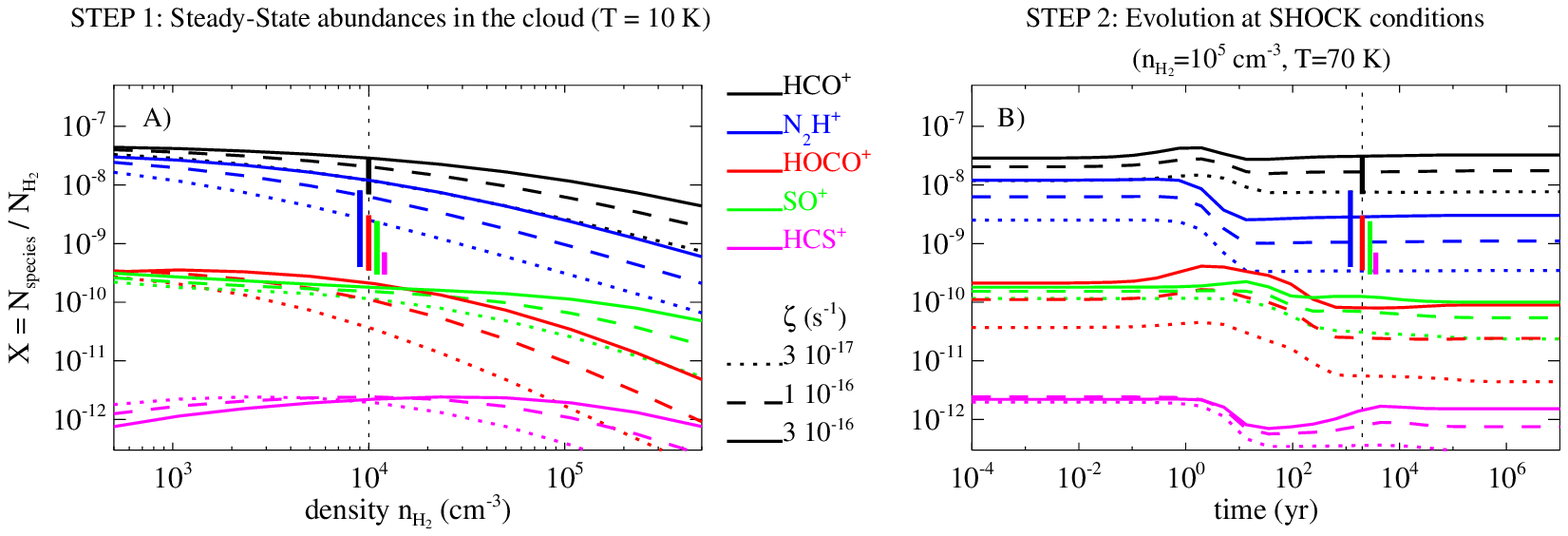}
   \caption{Observed abundances of molecular ions (colour vertical bars) are compared with theoretical values calculated with Astrochem (colour lines). The colors indicate \hcop\ (black), \nnhp (blue), \hocop\ (red), \sop\ (green), and \hcsp\ (magenta) abundances. {\it Panel A)} ``STEP 1'': steady-state abundances in the cloud, X = N$_{\rm species}$/N$_{\rm H_2}$, as a function of the gas density, n$_{\rm H_2}$ (\cmc), for gas temperature T$_{\rm kin} =10$ K, and cosmic rays ionization rate $\zeta= 3 \ 10^{-17}, 10^{-16}, 3 \ 10^{-16}$ s$^{-1}$ (dotted, dashed, and solid lines). Observed abundances are overplot at the assumed cloud density (\denshh $= 10^4$ \cmc, vertical dotted line). {\it Panel B)} ``STEP 2'': evolution of steady-state abundances  in the shock as a function of time (in years). The gas density and temperature are raised from cloud values (\denshh $= 10^4$ \cmc, T$_{\rm kin} = 10$ K) to the observed values in the outflow cavity (\denshh $= 10^5$ \cmc, T$_{\rm kin} = 70$ K). Observed abundances are overplot at the shock age (t$_{\rm shock} \sim 2000$ years, vertical dotted line).}
   \label{fig:astrochem_step1}
    \end{figure*}

In order to understand the origin of the observed molecular ions we compare the abundances inferred from observations (see Table \ref{tab:lvg}) with predictions obtained via the chemical code Astrochem\footnote{http://smaret.github.io/astrochem/}.
Astrochem computes the chemical evolution of a gas at a given temperature, T$_{\rm kin}$, and hydrogen density, n$_{\rm H}$. The code reads a network of chemical reactions, builds a system of kinetic rates equations, and solves it using a state-of-the-art stiff ordinary differential equations (ODE) solver. It considers gas-phase processes and simple gas-grain interaction, such as freeze-out, and desorption via several mechanisms (thermal, cosmic-ray, and photo- desorption).
In our calculations we use the OSU\footnote{http://www.physics.ohio-state.edu/~eric/research.html} chemical network, we assume visual extinction A$_{\rm V} =10$ mag consistent with C$^{18}$O and $^{13}$CO column densities, a grain size of 0.1~\um, and initial elemental abundances as assumed by \citet{wakelam06b} in their low-metallicity model with the exception of He and N for which we assume cosmic abundances \citep[e.g., ][]{asplund05,tsamis11} (see Table \ref{tab:initial_abu}). 

\begin{table}
\caption[]{\label{tab:initial_abu} Initial elemental abundances with respect to \hh\ assumed in our chemical model.}
    \begin{tabular}[h]{cc}
\hline
Species & X = N$_{\rm species}$/\nhh \\
\hline
\hline
He      & 0.14 \\
N        & 7.40 10$^{-5}$ \\
O        & 3.52 10$^{-4}$ \\
C$^+$ & 1.46 10$^{-4}$ \\
S$^+$ & 1.60 10$^{-7}$ \\
Si$^+$ & 1.60 10$^{-8}$ \\
Fe$^+$ & 6.00 10$^{-9}$ \\
Na$^+$ & 4.00 10$^{-9}$ \\
Mg$^+$ & 1.40 10$^{-8}$ \\
\hline
    \end{tabular}
\end{table}

As a first step (``STEP 1''), we compute the evolution of the gas in the cloud before the passage of the shock, i.e. for a gas temperature T$_{\rm kin} = 10$ K.
Fig.~\ref{fig:astrochem_step1}A shows the steady-state abundances of the observed molecular ions for gas densities, \denshh, between $5 \ 10^2$ and $5 \ 10^5$ \cmc, and for three different values of the cosmic rays (CR) ionization rate ($\zeta = 3 \ 10^{-17}, 10^{-16}$, $3 \ 10^{-16}$ s$^{-1}$).
%Note that CR ionization rates higher than the typical interstellar value of $3 \ 10^{-17}$ s$^{-1}$ are expected in dense regions \citep{dalgarno06,padovani09}.
Note that the CR ionization rate in the interstellar medium varies between 10$^{-17}$ to a few 10$^{-16}$ s$^{-1}$ \citep{padovani09}. 
Steady-state abundances are reached on a timescale of $10^6 - 10^7$ years, i.e. much longer than the shock kinematical age (t$_{\rm shock} \sim 2000$ years, \citealt{gueth96}).
%The abundances estimated from observations are overplot at the typical dark cloud density, \denshh $= 10^4$ \cmc. 
%, where cosmic rays are splitted into lower energy but still ionizing ones 
The abundances of \hcop, \nnhp, and \hocop\ are enhanced by a factor of a few up to one order of magnitude for higher CR ionization rates, while they are decreased by 1 to 3 orders of magnitude with increasing density. 
%\hcop\ and \hcsp\ are also enhanced by a factor of a few with increasing temperature, while the abundance of the other ions is poorly dependent on T. 
The dependency of   \hcop, \nnhp, and \hocop\ abundance on $\zeta$ and \denshh\ is due to the fact that in the quiescent gas the main formation process of these ions is via reactions with \hhhp:
\begin{eqnarray}
{\rm H_3^+ + CO    \rightarrow HCO^+    + H_2} \\
{\rm H_3^+ + N_2   \rightarrow N_2H^+   + H_2} \\
{\rm H_3^+ + CO_2 \rightarrow HOCO^+ + H_2} \label{eq:hocop}
\end{eqnarray}
where \hhhp\ is produced by reaction of \hh\ with \hhp, which in turn is produced by cosmic ray ionization of \hh, and destructed by dissociative recombination with electrons. Hence the abundance of \hhhp\ and, consequently, of these molecular ions increases with $\zeta$ and decreases with density. 
S-bearing species, \sop\ and \hcsp, are less dependent on the assumed $\zeta$ and stay roughly constant for densities up to $10^5$ \cmc\ if $\zeta = 1 - 3 \ 10^{-16}$ s$^{-1}$. 

As a second step (``STEP 2''), we compute how the obtained steady-state abundances evolve when the gas is compressed and heated in the shock, i.e. when density and temperature raise from typical dark cloud values (\denshh $= 10^4$ \cmc, T$_{\rm kin} = 10$ K) to the observed values in the outflow cavity, $g_2$ (\denshh $= 10^5$ \cmc, T$_{\rm kin} = 70$ K).
%The evolution in the shock is calculated for the highest CR ionization rate, $\zeta = 3 \ 10^{-16}$ s$^{-1}$.
Fig.~\ref{fig:astrochem_step1}B shows that molecular ions chemistry proceeds fast, so  they adjust very rapidly to the change of physical conditions caused by the shock, on timescales shorter or comparable to the shock age ($\sim 2000$ years). The abundance of all molecular ions drops due to the higher density but this is partially compensated for by the temperature enhancement.  

\begin{figure}[!ht]
     \includegraphics[width=8.8cm]{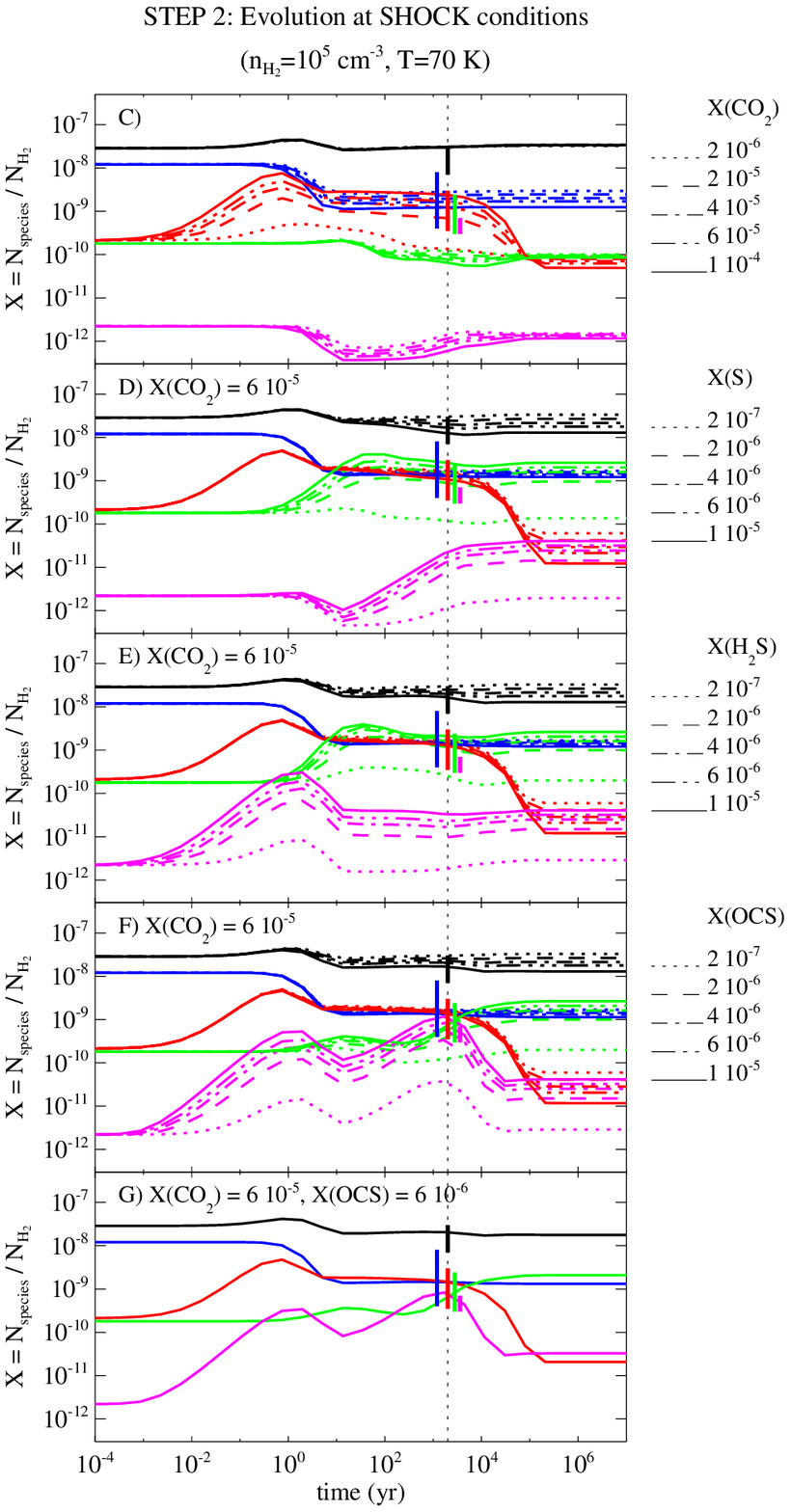}
   \caption{Evolution of molecular ions abundances in the shock as a function of time (``STEP 2''). The colors indicate \hcop\ (black), \nnhp (blue), \hocop\ (red), \sop\ (green), and \hcsp\ (magenta) abundances. Observed abundances (colour vertical bars) are overplot at the shock age (t$_{\rm shock} \sim 2000$ years, vertical dotted line). The evolution is computed from steady-state values in the cloud (\denshh $= 10^4$ \cmc, T$_{\rm kin} = 10$ K, $\zeta = 3 \ 10^{-16}$ s$^{-1}$) by enhancing the gas temperature and density (\denshh $= 10^5$ \cmc, T$_{\rm kin} = 70$ K), and the abundance of molecules which are thought to be sputtered off dust grain mantles: \coo, S, H$_2$S, and OCS ({\it panels C), D), E), and F)}, respectively). When varying S, H$_2$S, and OCS abundances, \xcoo\ is set to $= 6 \ 10^{-5}$. Finally, {\it panel G)} shows the evolution of molecular ions abundances in our ``best-fit model'', i.e. for \xcoo~$= 6 \ 10^{-5}$ and \xocs~$= 6 \ 10^{-6}$.}
   \label{fig:astrochem_step2}
    \end{figure}

Lastly, the evolution of molecular ions in the shock, i.e. ``STEP 2'', is re-calculated
%from steady-state abundances in our ``cloud model'' (\denshh $= 10^4$ \cmc, T = 10 K, $\zeta = 3 \ 10^{-16}$ s$^{-1}$) 
by enhancing (besides the gas temperature and density) the abundance of species which are thought to be sputtered off dust grain mantles, such as \coo\ and sulphur-bearing species (see Fig.~\ref{fig:astrochem_step2}).
This is done by assuming $\zeta = 3 \ 10^{-16}$ s$^{-1}$, which is the CR ionization value which best fit \hcop\ and \nnhp\ abundances (see below).

In the following, we discuss how the obtained steady-state abundances in the cloud and their evolution in the shock compare with observations for each of the detected molecular ions.
In Fig.~\ref{fig:astrochem_step1} and in Fig.~\ref{fig:astrochem_step2} the observed abundances are shown as colour bars covering the range of values inferred for the different components, or, when only one abundance value is estimated, the relative uncertainty (see Sect.~\ref{sect:phys}). 
Observations are overplot at the typical dark cloud density, \denshh $= 10^4$ \cmc, in Fig.~\ref{fig:astrochem_step1}A, and at the shock age, t$_{\rm shock} \sim 2000$ years, in Figs.~\ref{fig:astrochem_step1}B and \ref{fig:astrochem_step2}. \\

% IF WE WANT TO COMPARE ONLY WITH ABU ESTIMATED IN G2 CHANGE THE FIGURE: The line analysis performed in Sect.~\ref{sect:results} shows that the bulk of the emission in the detected molecular ions originate in the outflow cavities, thus we consider for each species the range of abundances estimated in the $g_2$ and $g_3$ components. 
% The line analysis performed in Sect.~\ref{sect:results} shows that the bulk of the emission in the detected molecular ions originate in the outflow cavity B1, i.e. in the  $g_2$ component. However, given the large uncertainty affecting the estimated abundances (around a factor three) the abundances shown in Fig.~\ref{fig:astrochem_step1} and \ref{fig:astrochem_step2} cover the range of abundances inferred for the different components. 

\textbf{ \textit{ \hcop\ and \nnhp\ abundances:}}\\
Fig.~\ref{fig:astrochem_step1}A shows that the observed \hcop\ and \nnhp\ abundances are in agreement with what predicted in the quiescent gas in the cloud at \denshh $= 10^4$ \cmc, T$_{\rm kin} = 10$ K.
In the shock, the abundance of \hcop\ remains roughly constant, as the decrease due to compression is compensated for by the raise in temperature, while \nnhp\ decreases by a factor five up to one order of magnitude for the lowest CR ionization rate (Fig.~\ref{fig:astrochem_step1}B). 
Hence, \hcop\ and \nnhp\ observed abundances  (\xhcop~$\sim0.7-3~10^{-8}$, \xnnhp~$\sim0.4-8~10^{-9}$) are simultaneously reproduced at the shock age only by assuming a high CR ionization rate, i.e. $\zeta = 1 - 3 \ 10^{-16}$ s$^{-1}$. 
Due to recombination, \hcop\ and \nnhp\ abundances are further decreased when neutral molecules are released off the dust grains. Thus, in the following we assume $\zeta = 3 \ 10^{-16}$ s$^{-1}$ to simultaneously fit the abundances of all the molecular ions (see below). 
The performed analysis indicates that, in agreement with what suggested by \citet{codella13}, \hcop\ and \nnhp\ are not enhanced in the shock but instead are a fossil record of the pre-shock gas.\\

%Panel a) of Figure \ref{fig:astrochem_abu} shows that if we set the density and temperature of the gas as inferred for the outflow cavity B1 (n$_{\rm H} = 10^5$ \cmc, T $= 70$ K) we are able to reproduce the observed \hcop\ and \nnhp\ abundances for $\zeta = 10^{-16}$.
%Moreover, the gas evolution shown in panel b) indicates that the steady state abundances are reached for t $\ge 10^5$ years, i.e. in a gas which has evolved for a time much longer than the shock kinematical age ($\tau_{shock} \sim 2000$ years, \citealt{gueth96}).
%We verified with Astrochem that molecular ions chemistry proceeds fast, so once steady-state abundances are reached, a density and temperature enhancement from typical dark cloud conditions (e.g. T $=10$ K, n$_{\rm H} = 10^4$ \cmc) to the observed values in the outflow cavity (T $=70$ K, n$_{\rm H} = 10^5$ \cmc) is followed by a variation of the abundances on a timescale of only a few tens years.
%Therefore, after the passage of the shock the \nnhp, \hcop\ abundances rapidly adjust to the temperature and density enhancement caused by the shock. 
%This is in agreement with a \nnhp\ decrease of 1--2 orders of magnitude in the shock as shown by Codella et al. (in print) using the UCL CHEM code. 
%Hence, the observed \hcop\ and \nnhp\ are not produced in the shock but instead are a fossil record of the pre-shock gas.\\

\textbf{ \textit{ \hocop\ abundance:}}\\
In contrast with \hcop\ and \nnhp, the predicted \hocop\ abundance in the cloud is lower than observed by a factor of a few up to two orders of magnitude depending on the assumed $\zeta$ (Fig.~\ref{fig:astrochem_step1}A).
Moreover, the \hocop\ abundance is further lowered following the gas compression in the shock, being one order of magnitude lower than observed even for the highest CR ionization rate $\zeta = 3 \ 10^{-16}$ s$^{-1}$  (Fig.~\ref{fig:astrochem_step1}B).
As mentioned above, \hocop\ is mainly formed via \coo\ reacting with \hhhp\ (see Eq.~\ref{eq:hocop}), hence it may be a useful probe of \coo, which lacks a permanent electric dipole moment making radio detection impossible.
A low \hocop\ abundance is expected in the cloud since most \coo\ is trapped on the icy dust mantles.
In effect, we find that the steady-state abundance of \coo\ in the cloud (\denshh $= 10^4$ \cmc, T$_{\rm kin} = 10$ K, $\zeta = 3 \ 10^{-16}$ s$^{-1}$)  is $3 \ 10^{-7}$, in agreement with what predicted by other chemical models \citep[e.g., ][]{lee96} and what measured toward various star-forming regions through observations of the 15 \um\ band \citep{vandishoeck96,dartois98a,boonman03}. 
%However \coo\ can be released in gas-phase by dust grains sputtering and/or evaporation in a shock. 
Based on \hocop\ observations in the Galactic center, \citet{minh91} first suggested that a high \hocop\ abundance may be caused by shocks, which release \coo\ molecules frozen on the mantles of dust grains into gas-phase via grain-grain (shattering) or gas-grain (sputtering) collisions and/or via evaporation by UV-photons in dissociative shocks. 
%This scenario is supported by the fact that the distribution of the \hocop\ 2$_{20}$-1$_{01}$ emission in the Galactic center is quite similar to that of $^{29}$SiO emission, which is commonly used as a shock tracer as Si is locked in the core of dust grains in the ISM \citep{deguchi06}.

We investigate this scenario with Astrochem by computing the evolution of \hocop\ in the shock when enhancing \coo\ with respect to its steady-state abundance. 
%(\xcoo = 2 $10^{-6}$, 2 $10^{-5}$, 4 $10^{-5}$, 6 $10^{-5}$, and 1 $10^{-4}$).
Note that we have direct evidence of dust grains sputtering in the B1 shock from SiO observations \citep{gueth98,nisini07}. 
Fig.~\ref{fig:astrochem_step2}C shows that \hocop\ is rapidly enhanced by more than one order of magnitude for increasing \xcoo\ and it stays high for $\sim 10^4$ years.
The observed value, \xhocop~$\sim 10^{-9}$, is matched at the shock age (t$_{\rm shock} \sim 2000$ years) for \xcoo~$\ge 2 \ 10^{-5}$, i.e. around two orders of magnitude larger than in the cloud.
This value is consistent with previous estimates of the amount of solid \coo\ trapped into the icy mantle of dust grains \citep[e.g., ][]{dhendecourt89,tielens91,degraauw96b}.
Thus, this result may indicate total evaporation of the ice mantles, in agreement with the water abundance estimated by \citet{busquet14} in the hot shock component. 

The abundance of the other molecular ions is not (for \hcop) or only slightly (by a factor lower than two for the other molecular ions) affected by the \coo\ enhancement in the shock.\\

\textbf{ \textit{ \sop\ abundance:}}\\
In agreement with what estimated by previous gas-phase chemistry models \citep[e.g., ][]{herbst86}, our chemical model predicts an \sop\ steady-state abundance in the cloud of $1.8 \ 10^{-10}$, which is lowered down to $1 \ 10^{-10}$ in the shock. 
This is lower than observed by around one order of magnitude.
On the other hand, \citet{neufeld89a} have shown that the abundance of \sop\ is strongly enhanced in shocks due to the release of sulphur from dust mantles and its subsequent ionization, as \sop\ is primarly formed via \splus\ + OH $\rightarrow$ \sop\ + H. 

We explore this scenario by computing the evolution of molecular ions in the shock when enhancing the abundance of atomic S or H$_2$S (Figs.~\ref{fig:astrochem_step2}D and \ref{fig:astrochem_step2}E, respectively).
%from SS abundances as obtained from our ``cloud model''
Note that S and H$_2$S are enhanced up to values which are larger than the initial abundance of S$^+$ in the model, as most of S is assumed to be trapped onto dust grains initially.
However, they are still consistent with the solar abundance of S \citep[e.g., ][]{asplund05}. 
Following our analysis of \hocop\ we also set \xcoo~$= 6 \ 10^{-5}$.
Fig.~\ref{fig:astrochem_step2} shows that, if the released S or H$_2$S are $\sim 2 \ 10^{-6}$ and $\ge 2 \ 10^{-7}$ respectively, \sop\ is rapidly (less than 10 years) enhanced behind the shock and its abundance remains high over large timescales, matching the observed value (\xsop~$\sim 8 \ 10^{-10}$) at the shock age.  
This confirms a shock origin for \sop.

%Note that once H$_2$S is released off the dust grains its abundance decreases from $\ge 2 \ 10^{-6}$ down to $\sim 9 \ 10^{-7}$ at t$\sim 2000$ years, and $\sim 3 \ 10^{-7}$ at t$\sim 5000$ years.
%Within the uncertainties this can be considered in agreement with the H$_2$S abundance estimated  by \citet{bachiller97} (\xhhs~$\sim 3 \ 10^{-7}$). 
Note that the amount of H$_2$S which needs to be released in gas-phase to reproduce the observed \sop\ abundance is in agreement with the H$_2$S abundance in the B1 shock estimated by \citet{bachiller97} (\xhhs~$\sim 3 \ 10^{-7}$). 

Concerning the other molecular ions we note that the abundances of \nnhp\ and \hocop\ are poorly affected by S or H$_2$S enhancement, while the abundance of \hcop\ is lowered by a factor lower than two, still matching the observed value.
\hcsp\ is also enhanced but not enough to match the observed abundance (see below). \\

%Panel c2) shows the evolution of molecular ions abundances computed with Astrochem.
%The \sop\ abundance is efficiently enhanced up to values of $10^{-7}$ on a 10--100 years scale, then slowly decreases matching the observed value at the shock age (i.e. for $t = 2000$ years).
%This confirms a shock origin for \sop.
%Moreover, we explored, as an alternative scenario, the evolution of X(\sop) for enhanced S or H$_2$S abundances and find that in order to reproduce the observed \sop\ abundance we need not only a release of S in gas-phase but a strong ionization to produce S$^+$.
%This suggests that non-dissociative shocks are not producing \sop\ because they are not enough energetic to produce \splus. Thus, the detection of \sop\ in a shocked environment is a unique probe of the dissociative nature of the shock.

%Note that the abundance of the other molecular ions is poorly affected by the \splus\ enhancement, with the exception of \hcop\ whose abundance is increased up to $10^{-7}$, which is even higher than the observed value.        \\

\textbf{ \textit{ \hcsp\ abundance:}}\\
Fig.~\ref{fig:astrochem_step1} shows that the \hcsp\ abundance in the cloud is very low $\sim 10^{-12}$, even for high CR ionization rates, and it remains roughly unaltered after the passage of the shock. % for the selected $\zeta = 3 \ 10^{-16}$ s$^{-1}$.
The abundance of \hcsp\ is strongly related to that of CS, as this ion is formed when CS reacts with \hcop, \hhhp, and H$_3$O$^{+}$, while the main destructive mechanism is dissociative recombination into CS.
Hence, the low values of \xhcsp\ in the cloud are likely due to a lack of CS.
Indeed, the steady-state abundance of CS in the cloud (\denshh $= 10^4$ \cmc, T$_{\rm kin} = 10$ K, $\zeta = 3 \ 10^{-16}$ s$^{-1}$) is of $\sim 10^{-10}$, which is three orders of magnitude lower than what estimated in the B1 shock by \citet{bachiller97} (\xcs~$\sim 2 \ 10^{-7}$). 
%This hypothesis is supported by the finding that existing gas-phase chemical models are not able to reproduce the observed CS and \hcsp\ abundances \citep[e.g., ][]{doty04,wakelam04}.
%This is observed also in the case of L1157-B1 where the inferred CS abundance (\xcs~$\sim 2-0.9~10^{-7}$, \citealt{bachiller97}, Gomez-Ruiz et al. in prep) is two orders of magnitude larger than what calculated with Astrochem. 

Figs.~\ref{fig:astrochem_step2}D and \ref{fig:astrochem_step2}E show that \hcsp\ is enhanced by almost two orders of magnitude when S or H$_2$S are released off the dust grains but is still lower than observed by one order of magnitude.
However, larger CS and \hcsp\ abundances can be obtained if we assume that OCS is one of the main sulphur carrier on dust grains, as suggested by \citet{wakelam04,wakelam05} and \citet{codella05}. 
In particular, Fig.~\ref{fig:astrochem_step2}F shows that the observed \hcsp\ abundance (\xhcsp~$\sim 3 - 7 \ 10^{-10}$) is matched at the shock age when the abundance of OCS in the shock is enhanced up to values $\ge 2 \ 10^{-6}$. 
Interestingly, also \xcs\ turns out to be in agreement with the observed value within a factor two (\xcs~$\sim 2 \ 10^{-7}$, \citealt{bachiller97}).

On the other hand we note that if the amount of OCS released off the dust grains is $\ge 2 \ 10^{-6}$ the OCS abundance at the shock age is $\sim 10^{-6}$, and it decreases down to value of $\sim 10^{-7}$ only on timescales $\ge 10^4$ years.
Thus, \xocs\ at the shock age is around 1--2 orders of magnitude larger than what estimated in the B1--B2 cavities by \citet{bachiller97} (\xocs~$\sim 3 \ 10^{-8} - 2 \ 10^{-7}$). 
%However, the estimate of \xocs\ obtained by \citet{bachiller97} may be affected by large uncertainty as it is derived using only three lines (9--8, 12--11, 19--18) and assuming a single excitation temperature (T $=70$ K) despite the large difference in the upper level energies of the considered lines (E$_{\rm up} = 26, 46, 111$ K). Our IRAM-30m/{\it Herschel}/HIFI survey reveals a large number of OCS lines with energies up to $200$ K, which would require a multiple gas components LVG analysis. This, however, is beyond the scope of the present paper.
In this regard, several authors point out that at present chemical models are unable to explain simultaneously CS, \hcsp, and OCS abundances and that the chemistry of sulphur-bearing species is not fully understood \citep{doty04,wakelam04,wakelam05}.
\citet{doty04} note that  the rates of ion-molecule reaction producing \hcsp, i.e. (\hcop, \hhhp, H$_3$O$^{+}$) + CS, have been only estimated but would need to be confirmed by laboratory experiments.\\

\textbf{ \textit{ ``Best-fit model'':}}\\
Fig.~\ref{fig:astrochem_step2} shows that the abundances of the detected molecular ions (see Tab.~\ref{tab:lvg}) are well reproduced by the gas evolution in the shock if \coo, S, and OCS are substantially enhanced due to dust grains sputtering and/or evaporation.
In particular, we are able to simultaneously match all the estimated abundances at the shock age ($\sim 2000$ years) if we assume \xcoo~$= 6 \ 10^{-5}$ and \xocs~$= 6 \ 10^{-6}$  (see Fig.~\ref{fig:astrochem_step2}G).
In order to verify the robustness of our ``best-fit model'', we verify that the abundances of other species, both ionic and neutral, are consistent with present and past observations.
%we use it to compute the evolution of all species in the chemical network.
The abundances obtained at the shock age of molecular ions and anions which are not-detected in our survey are summarized in Tab.~\ref{tab:abu_ions}.
Note that they are all $< 10^{-10}$, hence consistent with non-detection of the transitions falling in the spectral range covered by our survey, with the exception of H$_3$O$^{+}$, whose predicted abundance is $8 \ 10^{-9}$.
The only H$_3$O$^{+}$ transition covered by our survey is the transition at 307.2 GHz, observed with WILMA with a resolution of 2~MHz, for which we derive an upper limit of 0.03 K \kms.
This is consistent with non-detection when assuming the gas temperature and density inferred for the $g_2$ component. 
%The only H$_3$O$^{+}$ transition covered by our survey is the transition at 307.2 GHz, which is observed only with WILMA, i.e. with a resolution of 2~MHz. 
%The line is not detected  down to an rms of 3.5 mK per interval of 2 MHz, which implies an upper limit on the H$_3$O$^+$ abundance of {\bf ... to calculate}, which is in agreement {\bf or not} with the abundance predicted by our best-fit model.\\

%{\bf PROBLEMS and TO-DOs:\\ 

%\begin{itemize}

%\item[1)] \xocs\ in B1 is only $3 \ 10^{-8}$ \citep{bachiller97}, while in our ``best-fit model'' is $3.5 \ 10^{-6}$ at t$_{\rm shock} = 2000$ years, and $10^{-9}$ at SS.\\
%CECILIA: check the OCS evolution in our ``best-fit model'' to see if it raplidy decreases on a timescale comparable to the shock age? \\
%BERTRAND:  is looking at OCS lines to derive an accurate estimate of \xocs\ in B1.\\

%\item[2)] BERTRAND: search for the  H$_3$O$^{+}$ transition at 307.2 GHz covered with WILMA at 2MHz resolution.\\

%\end{itemize}
%}

\begin{table}
\caption[]{\label{tab:abu_ions} Abundances of non-detected molecular ions with respect to \hh, as obtained from our ``best-fit model'' (see Fig.~\ref{fig:astrochem_step2}G).}
    \begin{tabular}[h]{cc}
\hline
Species & X = N$_{\rm species}$/\nhh \\
\hline
\hline
              CO$^+$    & 6 10$^{-14}$ \\
             H$_2$O$^+$  & 7 10$^{-13}$ \\
            H$_3$O$^+$ &  8 10$^{-9}$ \\
              CH$^+$    & 6 10$^{-16}$ \\
             HOC$^+$   & 1 10$^{-11}$ \\
            HCNH$^+$ & 4 10$^{-11}$ \\
             NH$_4$$^+$   & 9 10$^{-11}$\\
               CN$^-$ & 2 10$^{-18}$\\
              C$_3$N$^-$ & 3 10$^{-18}$\\
              C$_4$H$^-$ & 1 10$^{-16}$\\
\hline
    \end{tabular}
\end{table}

\section{Conclusions}
\label{sect:conclusions}

In this paper we presented a comprehensive study of the chemistry of molecular ions in protostellar shocks.
As part of the ASAI and CHESS projects, we performed a census of molecular ions in the chemically rich protostellar shock L1157-B1 by means of an unbiased and high sensitivity ($\sim 1.5$~mK in the 3~mm band) spectral survey executed with the IRAM-30m telescope and {\it Herschel}/HIFI. %, down to a sensitivity of $\sim 1.5$~mK per 1.4 \kms\ interval in the 3~mm band.
%This allowed us to search for molecular ions emission down to a sensitivity of $\sim 1.5$~mK per 1.4 \kms\ interval in the 3~mm band.
%We analysed the observed lines using an LVG radiative transfer code and we inferred the gas physical conditions and the fractional abundances of the detected molecular ions.
%The latter have been compared with steady-state abundances in the cloud and their evolution in the shock computed with the chemical code Astrochem.
From our analysis we obtained the following results: \\

\begin{itemize}

\item[-] we report emission from \hcop, \nnhp, \hcsp\ and, for the first time in a protostellar shock, from \hocop\ and \sop.\\

\item[-] all the detected lines peak at blueshifted velocity, $\sim 0.5-3$ \kms\ with respect to systemic (V$_{\rm sys} = + 2.6$~\kms), and have a line width of $\sim 3-7$~\kms. A higher velocity component (V up to $- 40$~\kms) associated with the primary jet is detected in the \hcop~1--0 line.\\

\item[-] the lines profiles can be decomposed into different velocity components following an exponential intensity-velocity relation as found for the CO lines \citep{lefloch12}. The bulk of the emission is associated with the extended outflow cavities and originates from a gas with temperature T$_{\rm kin} \sim 20-70$ K, and density \denshh $\sim 10^5$ \cmc, in agreement with what found for the CO lines.\\

\item[-] inferred \hcop\ and \nnhp\ abundances (\xhcop~$\sim0.7-3~10^{-8}$, \xnnhp~$\sim0.4-8~10^{-9}$) are in agreement with on-source values \citep{bachiller97}, suggesting that these species are not enhanced in the shock. This is further confirmed by the fact that the observed abundances are in agreement with steady-state abundances in the cloud (i.e. at T$_{\rm kin} = 10$~K, \denshh~$= 10^4$ \cmc). The decrease of their abundance due to gas compression in the shock is compensated for by the raise in temperature, and predicted values match the observations for high cosmic rays ionization rate $\zeta = 3 \ 10^{-16}$ s$^{-1}$. Hence, \hcop\ and \nnhp\ are a fossil-record of the gas chemistry before the arrival of the shock.\\

\item[-] \hocop, \sop, and \hcsp\ abundances  (\xhocop~$\sim10^{-9}$, \xsop~$\sim8~10^{-10}$, \xhcsp~$\sim3-7~10^{-10}$), instead, are 1--2 orders of magnitude larger than predicted steady-state abundances in the cloud. 
On the other hand, they are strongly enhanced in the shock if \coo, S, H$_2$S, and OCS abundances are increased with respect to steady-state due to release off the dust grains in the shock. Hence, as suggested by previous studies \citep[e.g. ][]{minh91,deguchi06,turner92,turner94}, these species are effective shock tracers.\\

\item[-] the observed abundances allow constraining the amount of \coo\ and sulphur-bearing species released in the shock. 
To reproduce \hocop\ observations, \coo\ should be enhanced up to abundances $\ge 2 \ 10^{-5}$, i.e. two order of magnitude larger than its gas-phase abundance in the quiescent gas. This is in agreement with previous estimate of the \coo\ content on dust mantles.
To reproduce \sop, instead, the sulphur released in the form of H$_2$S molecules should be $\ge 2 \ 10^{-7}$. This value is in agreement with the H$_2$S abundance in B1 estimated by \citet{bachiller97}.\\

\item[-] as noted in previous studies, CS and \hcsp\ abundances in the quiescent gas in dark clouds are very low.  However, the observed \hcsp\ and CS abundances (\xhcsp~$\sim 3-7~10^{-10}$, \xcs~$\sim 2~10^{-7}$) are matched if  OCS is assumed to be one of the main sulphur career on dust grains, as suggested by \citet{wakelam04,wakelam05} and \citet{codella05}. Hence its abundance is enhanced to values $\ge 2 \ 10^{-6}$ in the shock. 
Such high \xocs\ values, however, are at least one order of magnitude larger than estimated by \citet{bachiller97} and are lowered to the observed values only on timescales $\ge 10^4$ years, i.e. much larger than the shock age (t$_{\rm shock} \sim 2000$ years).  
This difficulty in simultaneously reproducing CS, \hcsp, and OCS abundances was already reported by other authors and indicates that the chemistry of sulphur-bearing species is not yet fully understood \citep{doty04,wakelam04,wakelam05}. 
In this regard, laboratory experiments are required to verify the rates of ion-molecule reaction producing \hcsp, i.e. (\hcop, \hhhp, H$_3$O$^{+}$) + CS. \\

%\item[-] the abundances of all the detected molecular ions are simultaneously reproduced by assuming \xcoo~$= 6 \ 10^{-5}$ and \xocs~$= 6 \ 10^{-6}$ in the shock.
%The abundance of non-detected molecular ions and anions predicted by this model is consistent with non-detection.\\

\end{itemize}

\begin{acknowledgements}
L. Podio acknowledges the funding from the FP7 Intra-European Marie Curie Fellowship (PIEF-GA-2009-253896).
B. Lefloch and C. Ceccarelli acknowledge the financial support by the French space agency CNES.
R. Bachiller acknowledges partial support from Spanish MINECO under grant FIS2012-32096.
\end{acknowledgements}

\bibliographystyle{aa} % style aa.bst 
%\bibliography{../mybibtex} % your references Yourfile.bib

\end{document}